\documentclass[aps,prd,nofootinbib,amsmath,twocolumn]
{revtex4}
\usepackage{amssymb,esvect,amsmath,graphicx,latexsym,amsthm,slashed,eso-pic,amsfonts,hyperref}
\usepackage[final]{pdfpages}
\newcommand{\be}{\begin{equation}}
\newcommand{\ee}{\end{equation}}
\newcommand{\nl}{\nonumber \\}

\usepackage{verbatim}

\newcommand{\TeV}{\text{ TeV}}
\newcommand{\GeV}{\text{ GeV}}
\newcommand{\MeV}{\text{ MeV}}
\newcommand{\keV}{\text{ keV}}
\newcommand{\eV}{\text{ eV}}
\newcommand{\mpl}{m_\mathrm{Pl}}

\newcommand{\x}{\chi}
\newcommand{\p}{\varphi}

\newcommand{\mdm}{m_{_\text{DM}}}

\newcommand{\Neff}{N_\text{eff}}

\newcommand{\Tg}{T} 
\newcommand{\Tke}{T^{\x \, \text{eq}}} 
\newcommand{\Tdec}{T^{\x \, \text{dec}}} 
\newcommand{\Tnu}{T^{\nu \, \text{dec}}} 
\newcommand{\Tbbn}{T^\text{BBN}} 
\newcommand{\gstarDM}{g_*^\x} 
\newcommand{\gstarnu}{g_*^\nu} 
\newcommand{\xiDMinitial}{\xi_\x^0} 
\newcommand{\xiDMinitialFourth}{\xi_\x^{0 \, 4}} 
\newcommand{\xiDM}{\xi_\x} 
\newcommand{\xinuinitial}{\xi_\nu^{\mathrm{SM}}} 
\newcommand{\xinu}{\xi_\nu} 

\def\lsim{\mathrel{\raise.3ex\hbox{$<$\kern-.75em\lower1ex\hbox{$\sim$}}}}
\def\gsim{\mathrel{\raise.3ex\hbox{$>$\kern-.75em\lower1ex\hbox{$\sim$}}}}

\newcommand{\order}[1]{\mathcal{O}{(#1)}}

\begin{document}

\hspace{13cm} \parbox{5cm}{SLAC-PUB-16989}~\\

\hspace{13cm}

\title{Thermal Dark Matter Below an MeV}
\author{Asher Berlin, Nikita Blinov}

\affiliation{SLAC National Accelerator Laboratory, 2575 Sand Hill Road, Menlo Park, CA, 94025, USA}

\date{\today}

\begin{abstract}

We consider a class of models in which thermal dark matter is lighter than an MeV. If dark matter thermalizes with the Standard Model below the temperature of neutrino-photon decoupling, equilibration and freeze-out cools and heats the Standard Model bath comparably, alleviating constraints from measurements of the effective number of neutrino species. We demonstrate this mechanism in a model consisting of fermionic dark matter coupled to a light scalar mediator. Thermal dark matter can be as light as a few keV, while remaining compatible with existing cosmological and astrophysical observations. This framework motivates new experiments in the direct search for sub-MeV thermal dark matter and light force carriers.

\end{abstract}

\maketitle

The expectation of new physics at the electroweak scale has motivated the search for dark matter (DM) in the form of weakly interacting massive particles (WIMPs) with mass $10 \GeV \lesssim \mdm \lesssim 1 \TeV$. However, no direct signs of WIMPs have been observed to date~\cite{Aad:2015zva,Khachatryan:2014rra,Tan:2016zwf,Akerib:2015rjg,Akerib:2016vxi,Agnese:2015nto}. From a bottom-up perspective, astrophysical observations constrain the mass of DM to lie anywhere between $10^{-31} \GeV$ and $10^{58} \GeV$. 
The lack of evidence for the WIMP paradigm thus leaves an overwhelmingly vast range of viable DM masses.
However, the WIMP is only a specific example of DM that is thermally generated from the Standard Model (SM) bath, and in such models the acceptable range of masses is significantly reduced. 

If DM acquires its abundance through thermal contact with the SM, perturbative unitarity of the theory dictates that $\mdm \lesssim \order{100} \TeV$~\cite{Griest:1989wd}. Although no similar theoretical inconsistencies arise for small masses, $\mdm \gtrsim \order{1} \MeV$ is often quoted as a robust lower bound on the mass of thermal DM. It has been argued that any sub-MeV relic that is in thermal contact with the SM below the temperature of neutrino decoupling necessarily leads to measurable deviations in the expansion rate of the universe~\cite{Ho:2012ug,Steigman:2013yua,Boehm:2013jpa,Nollett:2013pwa,Nollett:2014lwa,Steigman:2014uqa,Green:2017ybv,Serpico:2004nm}. During radiation domination, the expansion rate is determined by the number of relativistic degrees of freedom, or, equivalently, by the effective number of neutrino species, $\Neff$,  at low temperatures. The value of $\Neff$ is constrained by the successful predictions of Big Bang nucleosynthesis (BBN) and observations of the cosmic microwave background (CMB). Thermal DM lighter than an MeV can substantially alter the observed value, $\Neff \simeq 3$, by heating or cooling neutrinos relative to photons or by contributing directly to the energy density as dark radiation.

In this Letter, we propose a novel mechanism to generate sub-MeV thermal relics that can be applied to a large class of models. 
Our key observation is that if a light state \emph{enters} thermal equilibrium with the SM after neutrino-photon decoupling, then the constraints from measurements of $\Neff$ during BBN and recombination are significantly relaxed.\footnote{Delayed thermalization has been used to alleviate cosmological constraints on light hidden sectors~\cite{Bartlett:1990qq,Chacko:2003dt,Chacko:2004cz} and sterile neutrinos, e.g., in regards to the LSND anomaly~\cite{Dasgupta:2013zpn,Mirizzi:2014ama,Cherry:2014xra,Chu:2015ipa,Cherry:2016jol}.
} 
If this light state couples only to neutrinos, equilibration draws heat from the SM bath without changing $\Neff$.
Later, when the species decouples, the neutrino bath is heated. Alternatively, if the light state couples only to photons, equilibration and decoupling increases and lowers $\Neff$, respectively. In either case, modifications to $\Neff$ are reduced at late times.
Although this idea can be applied to any light thermal relic, we will focus on the production of DM. In this case, thermal DM as light as the warm DM limit, $\sim \text{keV}$, is possible. Our study strengthens the case for new technologies in the search for sub-MeV thermal DM, such as superconducting detectors~\cite{Hochberg:2015pha,Hochberg:2015fth,Schutz:2016tid,Knapen:2016cue}.

In the standard WIMP freeze-out paradigm, DM, denoted by $\x$, enters equilibrium with the SM at early times and acquires a large thermal abundance. As the universe cools below the DM mass, its number density decreases as long as it remains in chemical equilibrium with the SM. 
If annihilations into SM particles are responsible for maintaining chemical equilibrium, the rate is proportional to the thermally-averaged cross-section $\langle \sigma v \rangle \sim \alpha_\x^2 / m_\x^2$. Thermal DM lighter than a GeV necessarily requires the presence of a new light mediator~\cite{Hut:1977zn,Lee:1977ua}.
The DM and SM sectors can be equilibrated through mediator decays and inverse decays. The corresponding rate is given by $\Gamma_{\x \, \text{eq}} \sim \alpha_\x^\prime \, m_\x^2 / T$, where $m_\x / T$ is a relativistic time-dilation factor and we have assumed similar masses for the DM and the mediator.
Equilibrium is achieved when $\Gamma_{\x \, \text{eq}}$ overcomes the Hubble expansion rate, $H \sim T^2 /\mpl$, at a photon temperature of $T = \Tke$, where $\mpl$ is the Planck mass. 
A rephrasing of the standard WIMP-miracle estimate implies that $\x$ saturates the observed relic abundance for
\be
\label{eq:mx1}
m_\x \sim \left( \alpha_\x / \alpha_\x^\prime\right)^{1/3} \left( T_\text{eq} / \mpl \right)^{1/6} \Tke
~,
\ee
where $T_\text{eq} \simeq 0.8 \eV$ is the temperature at matter-radiation equality. Neutrinos decouple from the photon plasma at a temperature of $\Tnu \sim 2 \MeV$~\cite{Enqvist:1991gx}. Equation~(\ref{eq:mx1}) implies that DM enters thermal equilibrium after neutrino decoupling ($\Tke \lesssim \Tnu$) for $m_\x \lesssim \keV \times ( \alpha_\x  / 10^5 \, \alpha_\x^\prime )^{1/3}$. 
In this case, $\alpha_\x \gg \alpha_\x^\prime$ is required to evade warm DM limits, which exclude 
$m_\x$ below a few keV. This hierarchy of couplings is possible if DM annihilations are enhanced relative to the decays of the light mediator into SM particles, e.g., through resonant processes or if the mediator interacts feebly with the SM, as in models of secluded DM~\cite{Pospelov:2007mp}.

\begin{figure}[t]
\hspace{-0.5cm}
\includegraphics[width=0.5\textwidth]{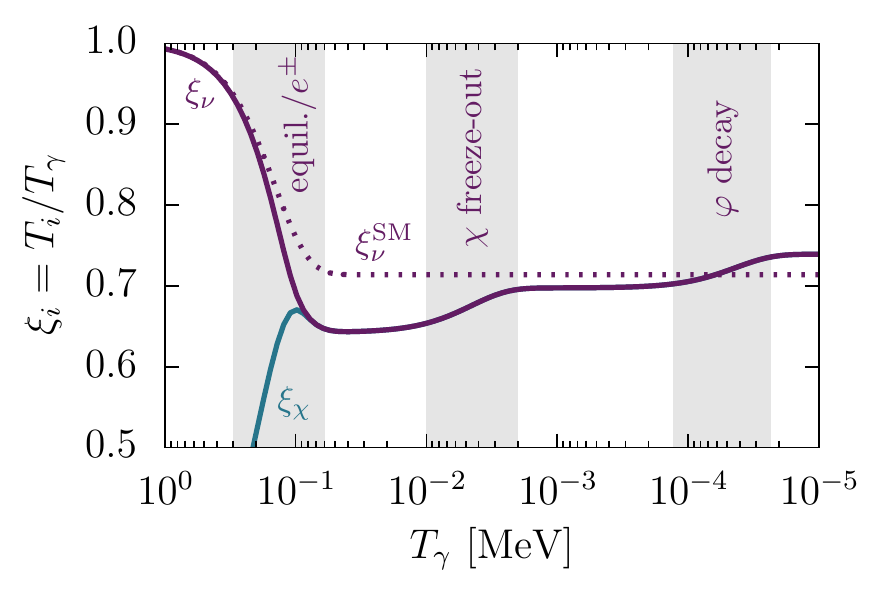} \hspace{-0.5cm}
\caption{Temperature evolution (normalized to the photon temperature) of the neutrino (upper, solid purple line) and DM (lower, solid cyan line) sectors in a model with a Majorana fermion, $\x$, coupled to a light real scalar mediator, $\p$. Note that the temperature of the photon bath decreases from left to right. Relative to the standard cosmology (dotted purple line), DM equilibration cools the SM neutrinos, while DM decoupling increases their temperature to a value close to the SM expectation. 
}
\label{fig:xi_evolution}
\end{figure}
Most investigations into the effects of additional light degrees of freedom assume that the new species is already in thermal equilibrium with the SM bath before neutrino-photon decoupling at $\Tnu \sim 2 \MeV$. We consider the complementary case in which a sub-MeV DM species, $\x$, enters thermal equilibrium with the SM at temperatures below $\Tnu$. New light particles that couple to electromagnetism are severely constrained by the anisotropies of the CMB and the observed cooling rates of stars and supernovae~\cite{Raffelt:1996wa,Ade:2015xua}. For simplicity, we will therefore focus on the case in which $\x$ couples exclusively to SM neutrinos. Even then, this new species can change the expansion history of the universe and thus may disrupt nucleosynthesis and recombination.

The formation of light nuclei occurs when the temperature of the photon bath falls in the range $50 \keV \lesssim T \lesssim 1 \MeV$. The presence of additional relativistic degrees of freedom during nucleosynthesis alters the predictions for the primordial abundances of light nuclei, such as Helium-4 and deuterium. For example, if $\Neff > 3$, neutrons freeze out earlier due to the increased expansion rate of the universe, resulting in a larger neutron to proton ratio at the onset of BBN and hence an enhanced Helium-4 yield. If the baryon density is fixed by the observed abundances of Helium-4 and deuterium, detailed considerations of BBN alone bound $\Neff \simeq 2.85 \pm 0.28$ during nucleosynthesis~\cite{Cyburt:2015mya}. Observations of the CMB angular power spectrum by the Planck satellite are also sensitive to the total radiation energy density at the time of recombination. Current measurements restrict the effective number of neutrino species at the time of photon decoupling with unprecedented precision, $\Neff \simeq 3.15 \pm 0.23$,
in impressive agreement with the standard $\Lambda \text{CDM}$ prediction of $\Neff \simeq 3.046$~\cite{Ade:2015xua}. Future CMB-S4 observations will be sensitive to deviations at the level of $\Delta \Neff \simeq \pm 0.027$~\cite{Abazajian:2016yjj}. The effects of a light 
dark sector on the expansion rate during these two well-studied epochs can therefore be encapsulated in the time evolution of $\Neff$.\footnote{Neutrino heating or cooling also directly affects the rate of the lingering neutron-to-proton conversion during BBN. We ignore this effect since we are most interested in times well after the freeze-out of weak interactions~\cite{Nollett:2013pwa,Nollett:2014lwa}.}

\begin{figure}[t]
\hspace{-0.5cm}
\includegraphics[width=0.5\textwidth]{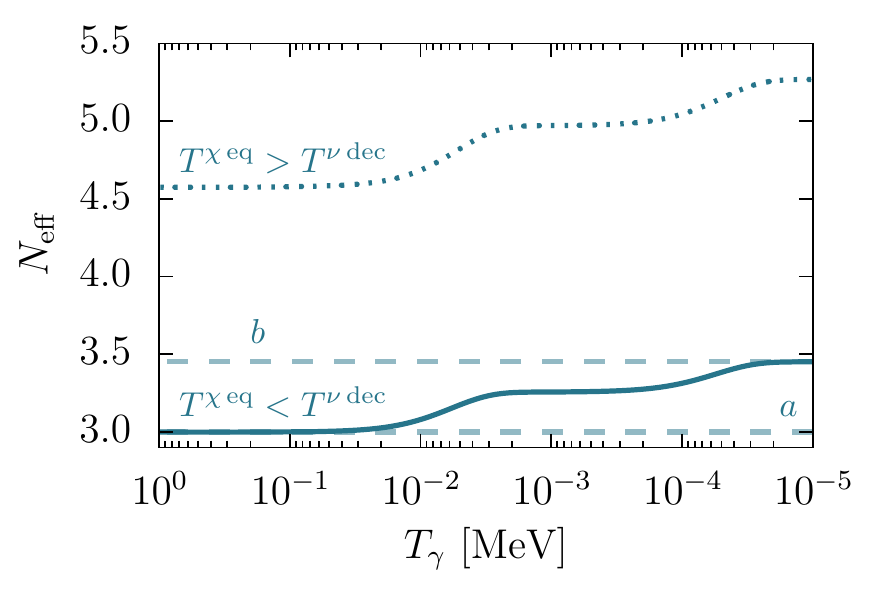} \hspace{-0.5cm}
\caption{Evolution of the effective number of neutrino species, $\Neff$, 
  for the scenario in Fig.~\ref{fig:xi_evolution} 
 (solid line). The dark sector equilibrates with the neutrinos after they decouple from the 
photon bath. The horizontal dashed lines correspond to the estimates in Eq.~(\ref{eq:Neff2}). 
The dotted line shows $\Neff$ if the dark sector equilibrates before 
neutrino decoupling.
}
\label{fig:Neff_and_Hubble_evolution}
\end{figure}

The temperature evolution of the neutrino and $\x$ populations can be derived from the conservation of the sum of their comoving energy or entropy densities. The effective number of relativistic degrees of freedom, $g_*^i$, in each population, $i = \nu , \x , \gamma$, determines the entropy density, $s_i \equiv (2 \pi^2 / 45) \, g_*^i \, T_i^3$, and energy density, $\rho_i \equiv (\pi^2/30) \, g_*^i \, T_i^4$, where $T_\gamma \equiv T$. Note that $\gstarDM$ includes contributions from additional particles in the DM sector. For three generations of SM neutrinos, we have $\gstarnu \equiv (7/8) \times 3 \times 2 = 21 /4$. 

If $\x$ equilibrates with the SM neutrinos below $\Tnu$ while relativistic, the energy density of the $\x$-neutrino bath, $\rho_{\nu + \x} \equiv \rho_\nu + \rho_\x$, scales as $a^{-4}$, where $a$ is the scale factor~\cite{Green:2017ybv}. This observation, together with 
entropy conservation of the photon bath, implies 
\be
(\gstarnu \, \xinu^4 + \gstarDM \, \xiDM^4 ) / (g_*^\gamma)^{4/3}  = \text{constant}
~
\label{eq:xi_relation_1}
\ee
before and immediately after equilibration, where $\xi_i \equiv T_i / \Tg$ is the temperature of species $i$ normalized to the photon temperature~\cite{Feng:2008mu}.  We also define $\xinuinitial$ as the value of $\xinu$ in the standard cosmology, such that $\xinuinitial (\Tg \gtrsim m_e) \equiv 1$, $\xinuinitial (\Tg \lesssim m_e) \equiv \left( 4/11 \right)^{1/3}$, in the instantaneous decoupling approximation. The value of $\xiDM$ before electron decoupling and DM-neutrino equilibration is denoted as $\xiDMinitial$ and encodes the UV-sensitivity of our model. 
Below $\Tke$, the entropy densities of the $\x$-neutrino bath, $s_{\nu + \x} \equiv s_\nu + s_\x$, and of the photons, $s_\gamma$, separately scale as $a^{-3}$. This implies that 
\be 
(\gstarnu \, \xinu^3 + \gstarDM \, \xiDM^3 ) / g_*^\gamma = \text{constant}
\label{eq:xi_relation_2}
\ee
before and after $\x$ becomes non-relativistic. Equations~(\ref{eq:xi_relation_1}) and~(\ref{eq:xi_relation_2}) can be used to determine the temperature evolution 
of the neutrinos and the dark sector. 
The effective number of neutrino species is then given by 
\be
\label{eq:Neff1}
\Neff \simeq 3 \, \bigg[ \bigg( \frac{\xinu}{\xinuinitial} \bigg)^4 + \Theta(T_\x - m_\x) ~ \frac{\gstarDM}{\gstarnu} ~ \bigg( \frac{\xiDM}{\xinuinitial} \bigg)^4 \, \bigg]
~,
\ee
where the step function encodes the instantaneous decoupling of $\x$ when its temperature drops below its mass, $T_\x \lesssim m_\x < \text{MeV}$~\cite{Feng:2008mu,Berlin:2016gtr}. Note that Eq.~(\ref{eq:Neff1}) reduces to $\Neff \simeq 3$ in the standard cosmology with $\gstarDM = 0$ and $\xinu = \xinuinitial$. 
We have approximated $3.046 \simeq 3$, neglecting the partial heating of neutrinos from $e^{\pm}$ annihilations~\cite{Mangano:2001iu,Mangano:2005cc}.

The contribution of the DM sector to the observed value of $\Neff$ depends on the ordering of the temperatures at which 
$\x$ enters and exits equilibrium with the SM ($\Tke \gg m_\x$ and $\Tdec \sim m_\x$, respectively) relative to $\Tnu \sim 2 \MeV$ and the temperature at which nucleosynthesis has effectively concluded, $\Tbbn \sim 50 \keV$. The possible orderings are
\begin{align}
\label{eq:scenarios}
& 1. ~~ \Tbbn < \Tnu < \Tke 
\nl
& 2. ~~ \Tbbn <  \Tdec < \Tke < \Tnu
\nl
& 3. ~~ \Tdec < \Tbbn <  \Tke < \Tnu
\nl
& 4. ~~ \Tdec <  \Tke < \Tbbn < \Tnu
~.
\end{align}
The first scenario above corresponds to standard weak-scale WIMPs, where $\x$ enters equilibrium with the SM before neutrino-photon decoupling. In the following three cases, $\x$ thermalizes with the SM neutrinos after they have decoupled from photons. The values of $\Neff$ in the latter scenarios are readily estimated using Eqs.~(\ref{eq:xi_relation_1})-(\ref{eq:Neff1}).
The effective number of neutrino species is given by
\be
\label{eq:Neff2}
\Neff \simeq
\begin{cases}
(a) ~ 3 \, \big( 1 + \frac{\gstarDM}{\gstarnu} ~ \xiDMinitialFourth \big) \\ 
(b) ~ 3 \left(1 + \frac{\gstarDM}{\gstarnu}\right)^{1/3} \left(1 + \frac{\gstarDM}{\gstarnu} \, \xiDMinitialFourth \right) ~, \\ 
\end{cases}
\ee
before/during and after $\x$ is in kinetic equilibrium with the neutrino bath, respectively. In scenario 2 of Eq.~(\ref{eq:scenarios}), $\x$ enters and exits kinetic equilibrium with the neutrino bath during nucleosynthesis and $\Neff$ evolves through (a) and (b) in Eq.~(\ref{eq:Neff2}) during BBN. In scenarios 3 and 4 of Eq.~(\ref{eq:scenarios}), $\x$ decouples after nucleosynthesis has concluded. In this case, $\Neff$ only takes the form of (a) in Eq.~(\ref{eq:Neff2}) during BBN. For every temperature ordering of Eq.~(\ref{eq:scenarios}), $\Neff$ is given by (b) in Eq.~(\ref{eq:Neff2}) at the time of recombination, relevant for measurements of the CMB. 
In Figs.~\ref{fig:xi_evolution} and \ref{fig:Neff_and_Hubble_evolution}, we show the temperature evolution of $\xi_{\nu, \x}$ and $\Neff$, respectively, for scenario 3 of Eq.~(\ref{eq:scenarios}) in the DM model described below. These results were obtained by solving Boltzmann equations for the neutrino and DM energy and entropy densities without relying on the instantaneous decoupling approximation. This numerical treatment agrees well with the approximate results of Eq.~(\ref{eq:Neff2}) and justifies their use in the remainder of this work.

\begin{figure}[t]
\hspace{-0.5cm}
\includegraphics[width=0.5\textwidth]{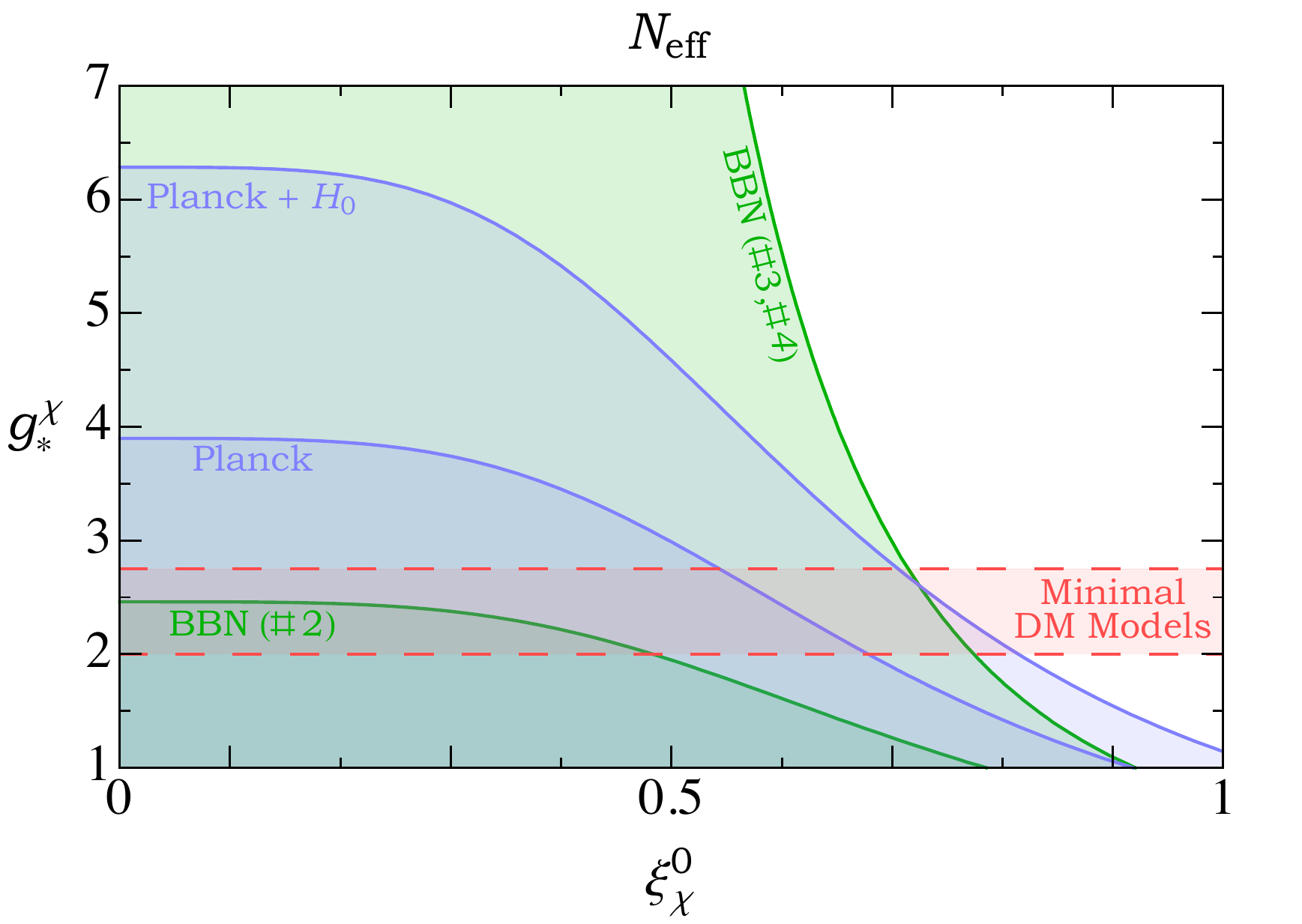} \hspace{-0.5cm}
\caption{Values of $g_*^\x$ (the effective number of sub-MeV dark sector states) and $\xiDMinitial$ (the initial 
dark sector-to-photon temperature ratio) compatible with the effective number of neutrino species at the time of nucleosynthesis (green) and recombination (blue) within $2 \sigma$. Contours labeled BBN ($\#2-\#4$) denote the temperature orderings of Eq.~(\ref{eq:scenarios}). The representative model space (red) corresponds to a dark sector with a DM scalar or Majorana fermion and a scalar mediator. 
}
\label{fig:Neff_Analytic}
\end{figure}

The analytic results of Eq.~(\ref{eq:Neff2}) show that for $\xiDMinitial \lesssim 1$, constraints on $\Neff$ are relaxed. If the DM sector is colder than the SM before equilibration ($\xi_\x^0 \ll 1$), then $\Neff \simeq 3 \left(1 + \gstarDM / \gstarnu \right)^{1/3}$ at late times. This is significantly reduced compared to the standard result, $\Neff \simeq 3 \left(1 + \gstarDM / \gstarnu \right)^{4/3}$, when $\Tke \gtrsim \Tnu$, as shown in Fig.~\ref{fig:Neff_and_Hubble_evolution}. Also in this limit, $\Neff \simeq 3$ while $\x$ is relativistic and in equilibrium with the neutrino bath, as required by conservation of comoving energy. Our scenario is different from the usual ``loophole'' where a cold dark sector does not equilibrate with the SM during DM freeze-out. If the initial $\x$ population is cooler than the SM plasma, then equilibration with the neutrino population drains neutrinos of thermal energy, lowering $\xinu$ below the standard prediction. This effect also increases $\xiDM$, which counteracts large changes to $\Neff$ (see Eq.~(\ref{eq:Neff1})). Once $\x$ decouples, conservation of comoving entropy demands that it heats the neutrino population close to its original temperature (relative to photons) for sufficiently small $\xiDMinitial$. This can be seen explicitly in Figs.~\ref{fig:xi_evolution} and \ref{fig:Neff_and_Hubble_evolution}. Note that this cyclic behavior of $\xi_\nu$ would not be possible if $\x$ equilibrated with the SM before neutrino-photon decoupling, as equilibration and decoupling would effectively borrow heat from the photon bath before pumping it into the neutrino sector.

  In Fig.~\ref{fig:Neff_Analytic}, we show the viable parameter space as a function of $\xiDMinitial$ and $\gstarDM$ for the different temperature orderings of Eq.~(\ref{eq:scenarios}). While a detailed simulation of nucleosynthesis is beyond the scope of this work, we conservatively demand that the maximum value of $|\Delta \Neff|$ obtained during BBN does not exceed the observed value by more than $2 \sigma$. The shaded regions denote parameters \emph{consistent} with observations, assuming that $\x$ equilibrates with the SM neutrinos after neutrino-photon decoupling. Also shown are regions consistent with $\Neff \simeq 3.15 \pm 0.23$ and $\Neff \simeq 3.3 \pm 0.3$, as derived from Planck data and its combination with local measurements of $H_0$~\cite{Ade:2015xua,Riess:2016jrr,Brust:2017nmv}. CMB-S4 observations will be sensitive to the entire parameter space shown.  

 In Fig.~\ref{fig:Neff_Analytic}, we also highlight representative model points, which correspond to a dark sector consisting of a DM scalar or Majorana fermion, $\x$, coupled to a real scalar mediator, $\p$, such that $\gstarDM = 2$ or $\gstarDM = (7/8) \times 2 + 1 = 2.75$, respectively. For concreteness, we will take our DM candidate, $\x$, to be a Majorana fermion. Note that if $\x$ was initially in thermal contact with the SM bath through the exchange of a heavy mediator, but decoupled above the electroweak scale, then conservation of comoving entropy dictates that $\xiDMinitial \simeq (10.75 / 106.75)^{1/3} \simeq 0.5$. This is a viable model of thermal DM if the mediator also couples to the SM neutrinos, $\nu$, 
allowing the dark sector to re-enter equilibrium with the SM below $\Tnu$.

The low-energy effective Lagrangian is given by
\be
\mathcal{L} \supset \p \, \left( \lambda_\x \, \x^2 + \lambda_\nu \, \nu^2 \right) + \text{h.c.}
~,
\ee
where $\chi$ and $\nu$ are two-component Weyl fermions and a sum over neutrino flavors is implied. We take $m_\x > m_\p$, such that $\x$ freezes out through annihilations into pairs of $\p$ particles, $\x \x \to \p \p$, after which $\p$ promptly decays into pairs of SM neutrinos.\footnote{Direct annihilations into neutrinos  ($\x \x \to \p \to \nu \nu$) are also viable, but require tuning $m_\p \sim 2 m_\x$. } The rate for the annihilation process is $p$-wave suppressed by the relative $\x$ velocity in the non-relativistic limit. Since this deposits a negligible amount of energy into the photon plasma at the time of recombination, there are no relevant limits derived from observations of CMB anisotropies~\cite{Ade:2015xua}. The neutrino-$\p$ coupling, $\lambda_\nu$, can arise from higher-scale interactions with right-handed neutrinos, $N$. For instance, in the standard seesaw mechanism, an interaction of the form $\p \, N^2$ naturally generates a SM neutrino coupling, $\lambda_\nu \sim m_\nu / m_N$, after electroweak symmetry breaking~\cite{Minkowski:1977sc,Yanagida:1979as,Mohapatra:1979ia,GellMann:1980vs,Schechter:1980gr}. Such interactions also arise in Majoron models of neutrino masses~\cite{Chikashige:1980ui,Gelmini:1980re,Georgi:1981pg,Aulakh:1982yn,Bertolini:1987kz,Babu:1991we}. 
Direct couplings to electrons are generated at one loop, but they are strongly suppressed by $\sim (g_2^2 m_e m_\nu^2) / (16 \pi^2 m_W^2 m_N )$.

In this model, various processes can establish kinetic equilibrium between the DM and SM sectors. Examples include elastic scattering between $\x$ and $\nu$ through $\p$ exchange, and decays and inverse-decays of $\p$ into pairs of SM neutrinos~\cite{Feng:2010zp}. 
We find that as long as $\x$ and $\p$ are chemically coupled, the latter process, $\p \leftrightarrow \nu \nu$, dominates kinetic equilibration between the DM and the neutrino bath since it is suppressed by fewer powers of small couplings. 

\begin{figure}[t]
\hspace{-0.5cm}
\includegraphics[width=0.5\textwidth]{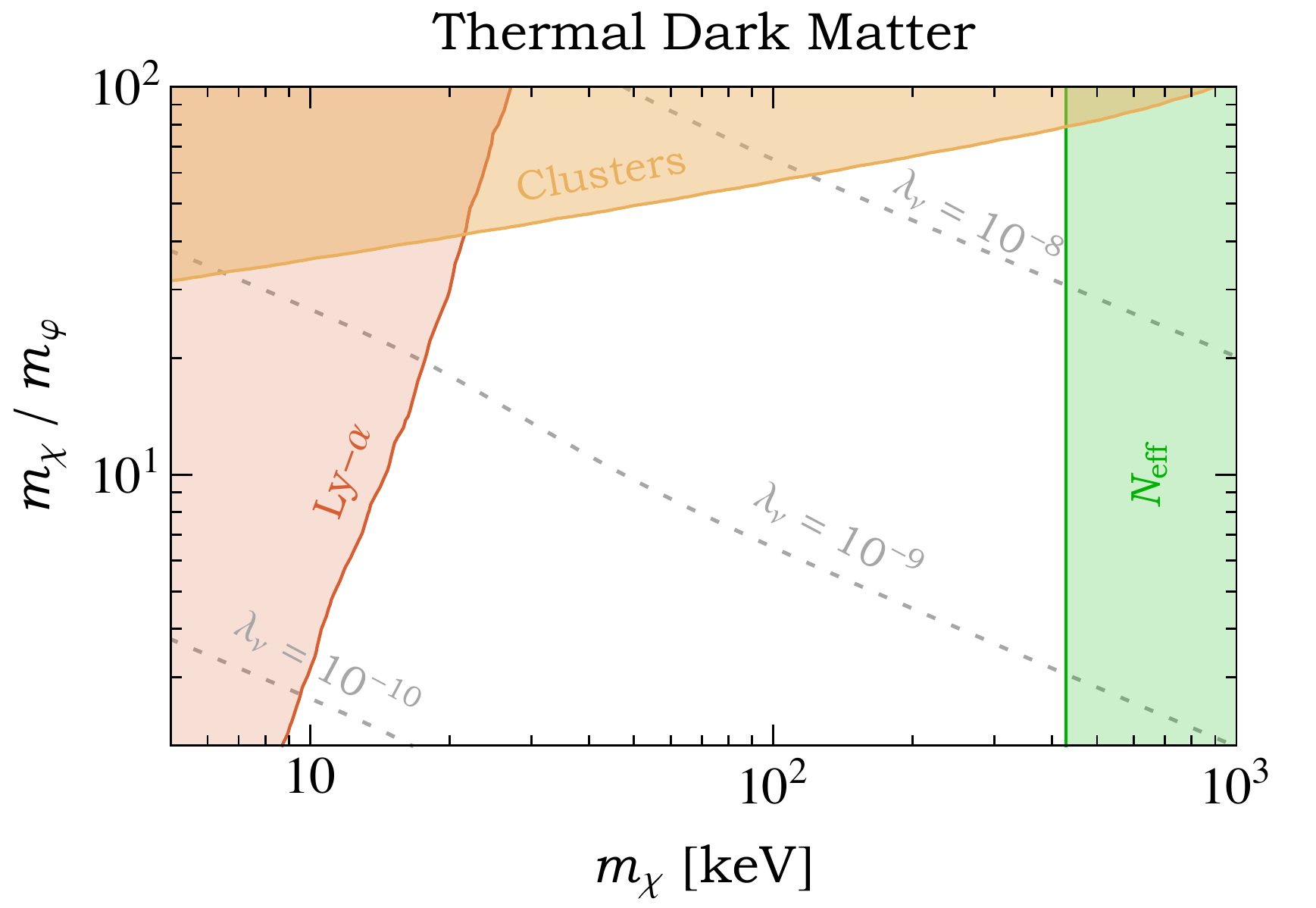} \hspace{-0.5cm}
\caption{The viable parameter space of our toy model. The DM-mediator coupling, $\lambda_\x$, is fixed to generate the correct abundance of $\x$, while the neutrino-mediator coupling, $\lambda_\nu$, (dotted grey) is chosen such that the DM sector thermalizes with SM neutrinos while relativistic. 
Measurements of $\Neff$ during nucleosynthesis and recombination exclude DM masses between several hundred keV and an MeV (green). Also shown are regions excluded by DM self-scattering in galaxy clusters (orange) and free-streaming of warm DM (red).}
\label{fig:Thermal}
\end{figure}

In Fig.~\ref{fig:Thermal}, we illustrate the viable parameter space of the representative toy model as a function of the DM mass, $m_\x$, and mass ratio, $m_\x / m_\p$, for $\xiDMinitial \lesssim 0.5$ (corresponding to the Planck bound in Fig.~\ref{fig:Neff_Analytic}). For each value of $m_\x$ and $m_\p$, the coupling $\lambda_\x$ is fixed to generate an abundance of $\x$ that is in agreement with the observed DM energy density. In the limit that $m_\x / m_\p \gg 1$, this requires $\lambda_\x \sim 10^{-5} \times (m_\x / \text{keV})^{1/2}$. We also demand that the DM sector equilibrates with the SM neutrinos while relativistic. This is accomplished by fixing $\lambda_\nu$ such that the rate for $\p \leftrightarrow \nu \nu$ overcomes the Hubble parameter at $\Tke \simeq 3 m_\x / \xiDM^\text{eq}$, corresponding to $\lambda_\nu \sim 10^{-11} \times (m_\x / m_\p) (m_\x / \text{keV})^{1/2}$. General constraints on $\Neff$ during nucleosynthesis and recombination are only relevant for $\Tke \gtrsim \Tnu$ and hence exclude sub-MeV DM masses greater than $\sim (2 \xiDM^\text{eq} / 3 ) \MeV \simeq 400 \keV$. As shown in previous studies, thermal DM heavier than several MeV is also viable as it is sufficiently non-relativistic at the time of neutrino decoupling~\cite{Ho:2012ug,Steigman:2013yua,Boehm:2013jpa,Nollett:2013pwa,Nollett:2014lwa,Steigman:2014uqa,Green:2017ybv,Serpico:2004nm}. For $m_\x \gtrsim 100 \keV$ and $m_\x / m_\p \lesssim 10$, $\x$ and $\p$ become non-relativistic at the tail-end of nucleosynthesis. Pending a detailed simulation of the light nuclei abundances, Fig.~\ref{fig:Neff_Analytic} suggests that these masses are possibly in slight disagreement with the predictions of BBN. If $\x$ is instead a real scalar boson, any possible tension is completely alleviated, while the general phenomenological picture in Fig.~\ref{fig:Thermal} remains qualitatively unchanged. We also note that while the parameter space shown is consistent with existing cosmological observations for 
$\xiDMinitial \lesssim 0.5$, it will be decisively tested by CMB-S4 experiments~\cite{Abazajian:2016yjj}.

Also shown in Fig.~\ref{fig:Thermal} are constraints from considerations of warm DM. Once $\x$ decouples from $\p$, it freely diffuses and suppresses matter perturbations below the free-streaming length, $\lambda_\text{fs}$. Limits derived from observations of the Lyman-$\alpha$ forest imply that $\lambda_\text{fs} \lesssim \order{1} \text{ Mpc}$~\cite{Viel:2013apy,Baur:2015jsy}. Such studies generally assume that DM has decoupled from the SM bath at early times, in which case $\lambda_\text{fs} \simeq 1.2 \text{ Mpc} \times (\mdm / \text{keV} )^{-1}$~\cite{Abazajian:2017tcc}. We translate the $2 \sigma$ lower bound on the warm DM mass from Ref.~\cite{Viel:2013apy}, $\mdm \gtrsim 3.3 \keV$, into the upper bound $\lambda_\text{fs} \lesssim 0.36 \text{ Mpc}$. We then compare this to the calculated value of the free-streaming length in our model, assuming that $\x$ begins diffusing once the rate for $\x \p \leftrightarrow \x \p$ falls below the Hubble expansion rate. Observations of the 21-cm hydrogen line during the cosmic dark ages could potentially improve the limits on $\lambda_\text{fs}$ and $m_\x$ by an order of magnitude~\cite{Sitwell:2013fpa,Sekiguchi:2014wfa,Shimabukuro:2014ava}.

DM self-scattering, $\x \x \to \x \x$, via $\p$ exchange, is constrained by studies of merging galaxy clusters. Following the discussion in Refs.~\cite{Tulin:2013teo,Tulin:2012wi}, we calculate the transfer scattering cross-section per DM mass, $\sigma_T / m_\x$, at the relevant virial velocities for galaxy clusters and demand that this does not exceed $1 \text{ cm}^2 / \text{g}$. As seen in Fig.~\ref{fig:Thermal}, this excludes values of $m_\x / m_\p \gtrsim \order{10}-\order{100}$ for $m_\x \sim \order{10} \keV - \order{100} \keV$, respectively. Additional cosmological and astrophysical constraints include modifications of the CMB from neutrino self-interactions~\cite{Cyr-Racine:2013jua,Lancaster:2017ksf}, diffusion-damped oscillations in the matter power spectrum from DM-neutrino scattering~\cite{Mangano:2006mp,Serra:2009uu,Wilkinson:2014ksa,Boehm:2000gq,Boehm:2004th,Boehm:2014vja,Schewtschenko:2015rno,Boehm:2003xr,Boehm:2001hm}, and supernovae cooling~\cite{Choi:1989hi,Kachelriess:2000qc,Farzan:2002wx,Heurtier:2016otg}. We find that these limits are negligible compared to the DM free-streaming and self-scattering constraints presented above. 

If the mediator-neutrino coupling is generated from interactions with right-handed sterile neutrinos, $N$, then at low-energies $\lambda_\nu \sim m_\nu / m_N$. Therefore, for SM neutrino masses at the level of $m_\nu \sim 0.1 \eV$, the values of $\lambda_\nu$ favored in Fig.~\ref{fig:Thermal} motivate sterile neutrinos near the GeV scale. It is intriguing to note that future direct searches for sterile neutrinos, such as the proposed SHiP experiment, could have sensitivity to couplings as small as $\lambda_\nu \sim 10^{-10}$~\cite{Alekhin:2015byh}. 

The minimal model presented above opens the cosmological window for sub-MeV thermal relics, but it does 
  not give rise to detectable signals at proposed low-threshold direct detection experiments~\cite{Hochberg:2015pha,Hochberg:2015fth,Schutz:2016tid,Knapen:2016cue}. However, simple scenarios that build upon our mechanism are within the projected reach of these technologies. For instance, interactions between DM and hadrons are present at low-energies
  if $\p$ also couples to a heavy generation of vector-like quarks, as in
  Ref.~\cite{Knapen:2017xzo}. As a result, $\p$ can be produced in supernovae, but leaves their cooling rates unchanged if  it becomes trapped, implying a \emph{lower bound} on the $\p$-nucleon coupling, $\lambda_n \gtrsim 10^{-7}$.
  In the early universe, DM can equilibrate with the neutrino or photon
  bath as described in our minimal model. 
  The nucleon coupling does not equilibrate the DM and SM sectors before neutrino-photon decoupling  
  provided that $\lambda_n \lesssim 10^{-5}$ and the reheat temperature of
  the universe 
  is $T_\text{RH} \sim 10 \MeV$.
  Furthermore, for nucleon couplings of this size and $m_\x \sim 100 \keV$, the
  DM-nucleon elastic scattering cross-section is $\sim 10^{-40} \text{ cm}^2$,
  which is well within the projected reach of proposed
  experiments~\cite{Battaglieri:2017aum}. 

In this Letter, we have considered a class of models in which sub-MeV thermal DM equilibrates with the SM neutrino bath after neutrino-photon decoupling. Dark sector equilibration and decoupling cools and reheats the neutrino sector close to its original temperature. Constraints derived from measurements of the effective number of neutrino species at the time of nucleosynthesis and recombination are significantly alleviated. While we have explicitly focused on DM, this mechanism could be more generally applied to any light thermal relics~\cite{Bartlett:1990qq,Chacko:2003dt,Chacko:2004cz}, such as those in Ref.~\cite{Beacom:2004yd}. Contrary to the standard lore, DM that acquires its abundance through thermal contact with the SM is viable for masses ranging from a keV to an MeV.

\section*{Acknowledgments}

We would like to thank Kevork Abazajian, Jonathan Feng, Alex Friedland, Anson Hook, Manoj Kaplinghat, Simon Knapen, Gordan Krnjaic, Tongyan Lin, Andrew Long, Gustavo Marques-Tavares, Brian Shuve, and Kathryn Zurek for valuable conversations. AB and NB are supported by the U.S. Department of Energy under Contract No. DE-AC02-76SF00515. 

\bibliography{submev}

\begin{thebibliography}{10}%
\makeatletter
\providecommand \@ifxundefined [1]{%
 \ifx #1\undefined \expandafter \@firstoftwo
 \else \expandafter \@secondoftwo
\fi
}%
\providecommand \@ifnum [1]{%
 \ifnum #1\expandafter \@firstoftwo
 \else \expandafter \@secondoftwo
\fi
}%
\providecommand \enquote [1]{``#1''}%
\providecommand \bibnamefont  [1]{#1}%
\providecommand \bibfnamefont [1]{#1}%
\providecommand \citenamefont [1]{#1}%
\providecommand\href[0]{\@sanitize\@href}%
\providecommand\@href[1]{\endgroup\@@startlink{#1}\endgroup\@@href}%
\providecommand\@@href[1]{#1\@@endlink}%
\providecommand \@sanitize [0]{\begingroup\catcode`\&12\catcode`\#12\relax}%
\@ifxundefined \pdfoutput {\@firstoftwo}{%
 \@ifnum{\z@=\pdfoutput}{\@firstoftwo}{\@secondoftwo}%
}{%
 \providecommand\@@startlink[1]{\leavevmode\special{html:<a href="#1">}}%
 \providecommand\@@endlink[0]{\special{html:</a>}}%
}{%
 \providecommand\@@startlink[1]{%
  \leavevmode
  \pdfstartlink
   attr{/Border[0 0 1 ]/H/I/C[0 1 1]}%
   user{/Subtype/Link/A<</Type/Action/S/URI/URI(#1)>>}%
  \relax
 }%
 \providecommand\@@endlink[0]{\pdfendlink}%
}%
\providecommand \url  [0]{\begingroup\@sanitize \@url }%
\providecommand \@url [1]{\endgroup\@href {#1}{\urlprefix}}%
\providecommand \urlprefix [0]{URL }%
\providecommand \Eprint[0]{\href }%
\@ifxundefined \urlstyle {%
  \providecommand \doi [1]{doi:\discretionary{}{}{}#1}%
}{%
  \providecommand \doi [0]{doi:\discretionary{}{}{}\begingroup
  \urlstyle{rm}\Url }%
}%
\providecommand \doibase [0]{http://dx.doi.org/}%
\providecommand \Doi[1]{\href{\doibase#1}}%
\providecommand \bibAnnote [3]{%
  \BibitemShut{#1}%
  \begin{quotation}\noindent
    \textsc{Key:}\ #2\\\textsc{Annotation:}\ #3%
  \end{quotation}%
}%
\providecommand \bibAnnoteFile [2]{%
  \IfFileExists{#2}{\bibAnnote {#1} {#2} {\input{#2}}}{}%
}%
\providecommand \typeout [0]{\immediate \write \m@ne }%
\providecommand \selectlanguage [0]{\@gobble}%
\providecommand \bibinfo [0]{\@secondoftwo}%
\providecommand \bibfield [0]{\@secondoftwo}%
\providecommand \translation [1]{[#1]}%
\providecommand \BibitemOpen[0]{}%
\providecommand \bibitemStop [0]{}%
\providecommand \bibitemNoStop [0]{.\EOS\space}%
\providecommand \EOS [0]{\spacefactor3000\relax}%
\providecommand \BibitemShut [1]{\csname bibitem#1\endcsname}%
\bibitem{Aad:2015zva}%
  \BibitemOpen
  \bibfield{author}{%
  \bibinfo {author} {\bibfnamefont{G.}~\bibnamefont{Aad}} \emph{et~al.}
  (\bibinfo {collaboration} {ATLAS}),\ }%
  \bibfield{journal}{%
  \Doi{10.1140/epjc/s10052-015-3517-3, 10.1140/epjc/s10052-015-3639-7}{\bibinfo
  {journal} {Eur. Phys. J.}}\ }%
  \textbf{\bibinfo {volume} {C75}},\ \bibinfo {pages} {299} (\bibinfo {year}
  {2015}),\ \bibinfo {note} {[Erratum: Eur. Phys. J.C75,no.9,408(2015)]},\
  \Eprint{http://arxiv.org/abs/1502.01518}{arXiv:1502.01518 [hep-ex]}%
  \bibAnnoteFile{NoStop}{Aad:2015zva}%
\bibitem{Khachatryan:2014rra}%
  \BibitemOpen
  \bibfield{author}{%
  \bibinfo {author} {\bibfnamefont{V.}~\bibnamefont{Khachatryan}} \emph{et~al.}
  (\bibinfo {collaboration} {CMS}),\ }%
  \bibfield{journal}{%
  \Doi{10.1140/epjc/s10052-015-3451-4}{\bibinfo {journal} {Eur. Phys. J.}}\ }%
  \textbf{\bibinfo {volume} {C75}},\ \bibinfo {pages} {235} (\bibinfo {year}
  {2015}),\ \Eprint{http://arxiv.org/abs/1408.3583}{arXiv:1408.3583 [hep-ex]}%
  \bibAnnoteFile{NoStop}{Khachatryan:2014rra}%
\bibitem{Tan:2016zwf}%
  \BibitemOpen
  \bibfield{author}{%
  \bibinfo {author} {\bibfnamefont{A.}~\bibnamefont{Tan}} \emph{et~al.}
  (\bibinfo {collaboration} {PandaX-II}),\ }%
  \bibfield{journal}{%
  \Doi{10.1103/PhysRevLett.117.121303}{\bibinfo {journal} {Phys. Rev. Lett.}}\
  }%
  \textbf{\bibinfo {volume} {117}},\ \bibinfo {pages} {121303} (\bibinfo {year}
  {2016}),\ \Eprint{http://arxiv.org/abs/1607.07400}{arXiv:1607.07400
  [hep-ex]}%
  \bibAnnoteFile{NoStop}{Tan:2016zwf}%
\bibitem{Akerib:2015rjg}%
  \BibitemOpen
  \bibfield{author}{%
  \bibinfo {author} {\bibfnamefont{D.~S.}\ \bibnamefont{Akerib}} \emph{et~al.}
  (\bibinfo {collaboration} {LUX}),\ }%
  \bibfield{journal}{%
  \Doi{10.1103/PhysRevLett.116.161301}{\bibinfo {journal} {Phys. Rev. Lett.}}\
  }%
  \textbf{\bibinfo {volume} {116}},\ \bibinfo {pages} {161301} (\bibinfo {year}
  {2016}),\ \Eprint{http://arxiv.org/abs/1512.03506}{arXiv:1512.03506
  [astro-ph.CO]}%
  \bibAnnoteFile{NoStop}{Akerib:2015rjg}%
\bibitem{Akerib:2016vxi}%
  \BibitemOpen
  \bibfield{author}{%
  \bibinfo {author} {\bibfnamefont{D.~S.}\ \bibnamefont{Akerib}} \emph{et~al.}
  (\bibinfo {collaboration} {LUX}),\ }%
  \bibfield{journal}{%
  \Doi{10.1103/PhysRevLett.118.021303}{\bibinfo {journal} {Phys. Rev. Lett.}}\
  }%
  \textbf{\bibinfo {volume} {118}},\ \bibinfo {pages} {021303} (\bibinfo {year}
  {2017}),\ \Eprint{http://arxiv.org/abs/1608.07648}{arXiv:1608.07648
  [astro-ph.CO]}%
  \bibAnnoteFile{NoStop}{Akerib:2016vxi}%
\bibitem{Agnese:2015nto}%
  \BibitemOpen
  \bibfield{author}{%
  \bibinfo {author} {\bibfnamefont{R.}~\bibnamefont{Agnese}} \emph{et~al.}
  (\bibinfo {collaboration} {SuperCDMS}),\ }%
  \bibfield{journal}{%
  \Doi{10.1103/PhysRevLett.116.071301}{\bibinfo {journal} {Phys. Rev. Lett.}}\
  }%
  \textbf{\bibinfo {volume} {116}},\ \bibinfo {pages} {071301} (\bibinfo {year}
  {2016}),\ \Eprint{http://arxiv.org/abs/1509.02448}{arXiv:1509.02448
  [astro-ph.CO]}%
  \bibAnnoteFile{NoStop}{Agnese:2015nto}%
\bibitem{Griest:1989wd}%
  \BibitemOpen
  \bibfield{author}{%
  \bibinfo {author} {\bibfnamefont{K.}~\bibnamefont{Griest}}\ and\ \bibinfo
  {author} {\bibfnamefont{M.}~\bibnamefont{Kamionkowski}},\ }%
  \bibfield{journal}{%
  \Doi{10.1103/PhysRevLett.64.615}{\bibinfo {journal} {Phys. Rev. Lett.}}\ }%
  \textbf{\bibinfo {volume} {64}},\ \bibinfo {pages} {615} (\bibinfo {year}
  {1990})%
  \bibAnnoteFile{NoStop}{Griest:1989wd}%
\bibitem{Ho:2012ug}%
  \BibitemOpen
  \bibfield{author}{%
  \bibinfo {author} {\bibfnamefont{C.~M.}\ \bibnamefont{Ho}}\ and\ \bibinfo
  {author} {\bibfnamefont{R.~J.}\ \bibnamefont{Scherrer}},\ }%
  \bibfield{journal}{%
  \Doi{10.1103/PhysRevD.87.023505}{\bibinfo {journal} {Phys. Rev.}}\ }%
  \textbf{\bibinfo {volume} {D87}},\ \bibinfo {pages} {023505} (\bibinfo {year}
  {2013}),\ \Eprint{http://arxiv.org/abs/1208.4347}{arXiv:1208.4347
  [astro-ph.CO]}%
  \bibAnnoteFile{NoStop}{Ho:2012ug}%
\bibitem{Steigman:2013yua}%
  \BibitemOpen
  \bibfield{author}{%
  \bibinfo {author} {\bibfnamefont{G.}~\bibnamefont{Steigman}},\ }%
  \bibfield{journal}{%
  \Doi{10.1103/PhysRevD.87.103517}{\bibinfo {journal} {Phys. Rev.}}\ }%
  \textbf{\bibinfo {volume} {D87}},\ \bibinfo {pages} {103517} (\bibinfo {year}
  {2013}),\ \Eprint{http://arxiv.org/abs/1303.0049}{arXiv:1303.0049
  [astro-ph.CO]}%
  \bibAnnoteFile{NoStop}{Steigman:2013yua}%
\bibitem{Boehm:2013jpa}%
  \BibitemOpen
  \bibfield{author}{%
  \bibinfo {author} {\bibfnamefont{C.}~\bibnamefont{Boehm}}, \bibinfo {author}
  {\bibfnamefont{M.~J.}\ \bibnamefont{Dolan}},\ and\ \bibinfo {author}
  {\bibfnamefont{C.}~\bibnamefont{McCabe}},\ }%
  \bibfield{journal}{%
  \Doi{10.1088/1475-7516/2013/08/041}{\bibinfo {journal} {JCAP}}\ }%
  \textbf{\bibinfo {volume} {1308}},\ \bibinfo {pages} {041} (\bibinfo {year}
  {2013}),\ \Eprint{http://arxiv.org/abs/1303.6270}{arXiv:1303.6270 [hep-ph]}%
  \bibAnnoteFile{NoStop}{Boehm:2013jpa}%
\bibitem{Nollett:2013pwa}%
  \BibitemOpen
  \bibfield{author}{%
  \bibinfo {author} {\bibfnamefont{K.~M.}\ \bibnamefont{Nollett}}\ and\
  \bibinfo {author} {\bibfnamefont{G.}~\bibnamefont{Steigman}},\ }%
  \bibfield{journal}{%
  \Doi{10.1103/PhysRevD.89.083508}{\bibinfo {journal} {Phys. Rev.}}\ }%
  \textbf{\bibinfo {volume} {D89}},\ \bibinfo {pages} {083508} (\bibinfo {year}
  {2014}),\ \Eprint{http://arxiv.org/abs/1312.5725}{arXiv:1312.5725
  [astro-ph.CO]}%
  \bibAnnoteFile{NoStop}{Nollett:2013pwa}%
\bibitem{Nollett:2014lwa}%
  \BibitemOpen
  \bibfield{author}{%
  \bibinfo {author} {\bibfnamefont{K.~M.}\ \bibnamefont{Nollett}}\ and\
  \bibinfo {author} {\bibfnamefont{G.}~\bibnamefont{Steigman}},\ }%
  \bibfield{journal}{%
  \Doi{10.1103/PhysRevD.91.083505}{\bibinfo {journal} {Phys. Rev.}}\ }%
  \textbf{\bibinfo {volume} {D91}},\ \bibinfo {pages} {083505} (\bibinfo {year}
  {2015}),\ \Eprint{http://arxiv.org/abs/1411.6005}{arXiv:1411.6005
  [astro-ph.CO]}%
  \bibAnnoteFile{NoStop}{Nollett:2014lwa}%
\bibitem{Steigman:2014uqa}%
  \BibitemOpen
  \bibfield{author}{%
  \bibinfo {author} {\bibfnamefont{G.}~\bibnamefont{Steigman}}\ and\ \bibinfo
  {author} {\bibfnamefont{K.~M.}\ \bibnamefont{Nollett}},\ }%
  \bibfield{booktitle}{%
  \emph{\bibinfo {booktitle} {{Proceedings, 13th International Conference on
  Topics in Astroparticle and Underground Physics (TAUP 2013): Asilomar,
  California, September 8-13, 2013}}},\ }%
  \bibfield{journal}{%
  \Doi{10.1016/j.phpro.2014.12.029}{\bibinfo {journal} {Phys. Procedia}}\ }%
  \textbf{\bibinfo {volume} {61}},\ \bibinfo {pages} {179} (\bibinfo {year}
  {2015}),\ \Eprint{http://arxiv.org/abs/1402.5399}{arXiv:1402.5399
  [astro-ph.CO]}%
  \bibAnnoteFile{NoStop}{Steigman:2014uqa}%
\bibitem{Green:2017ybv}%
  \BibitemOpen
  \bibfield{author}{%
  \bibinfo {author} {\bibfnamefont{D.}~\bibnamefont{Green}}\ and\ \bibinfo
  {author} {\bibfnamefont{S.}~\bibnamefont{Rajendran}}}%
   (\bibinfo {year} {2017}),\
  \Eprint{http://arxiv.org/abs/1701.08750}{arXiv:1701.08750 [hep-ph]}%
  \bibAnnoteFile{NoStop}{Green:2017ybv}%
\bibitem{Serpico:2004nm}%
  \BibitemOpen
  \bibfield{author}{%
  \bibinfo {author} {\bibfnamefont{P.~D.}\ \bibnamefont{Serpico}}\ and\
  \bibinfo {author} {\bibfnamefont{G.~G.}\ \bibnamefont{Raffelt}},\ }%
  \bibfield{journal}{%
  \Doi{10.1103/PhysRevD.70.043526}{\bibinfo {journal} {Phys. Rev.}}\ }%
  \textbf{\bibinfo {volume} {D70}},\ \bibinfo {pages} {043526} (\bibinfo {year}
  {2004}),\
  \Eprint{http://arxiv.org/abs/astro-ph/0403417}{arXiv:astro-ph/0403417
  [astro-ph]}%
  \bibAnnoteFile{NoStop}{Serpico:2004nm}%
\bibitem{Bartlett:1990qq}%
  \BibitemOpen
  \bibfield{author}{%
  \bibinfo {author} {\bibfnamefont{J.~G.}\ \bibnamefont{Bartlett}}\ and\
  \bibinfo {author} {\bibfnamefont{L.~J.}\ \bibnamefont{Hall}},\ }%
  \bibfield{journal}{%
  \Doi{10.1103/PhysRevLett.66.541}{\bibinfo {journal} {Phys. Rev. Lett.}}\ }%
  \textbf{\bibinfo {volume} {66}},\ \bibinfo {pages} {541} (\bibinfo {year}
  {1991})%
  \bibAnnoteFile{NoStop}{Bartlett:1990qq}%
\bibitem{Chacko:2003dt}%
  \BibitemOpen
  \bibfield{author}{%
  \bibinfo {author} {\bibfnamefont{Z.}~\bibnamefont{Chacko}}, \bibinfo {author}
  {\bibfnamefont{L.~J.}\ \bibnamefont{Hall}}, \bibinfo {author}
  {\bibfnamefont{T.}~\bibnamefont{Okui}},\ and\ \bibinfo {author}
  {\bibfnamefont{S.~J.}\ \bibnamefont{Oliver}},\ }%
  \bibfield{journal}{%
  \Doi{10.1103/PhysRevD.70.085008}{\bibinfo {journal} {Phys. Rev.}}\ }%
  \textbf{\bibinfo {volume} {D70}},\ \bibinfo {pages} {085008} (\bibinfo {year}
  {2004}),\ \Eprint{http://arxiv.org/abs/hep-ph/0312267}{arXiv:hep-ph/0312267
  [hep-ph]}%
  \bibAnnoteFile{NoStop}{Chacko:2003dt}%
\bibitem{Chacko:2004cz}%
  \BibitemOpen
  \bibfield{author}{%
  \bibinfo {author} {\bibfnamefont{Z.}~\bibnamefont{Chacko}}, \bibinfo {author}
  {\bibfnamefont{L.~J.}\ \bibnamefont{Hall}}, \bibinfo {author}
  {\bibfnamefont{S.~J.}\ \bibnamefont{Oliver}},\ and\ \bibinfo {author}
  {\bibfnamefont{M.}~\bibnamefont{Perelstein}},\ }%
  \bibfield{journal}{%
  \Doi{10.1103/PhysRevLett.94.111801}{\bibinfo {journal} {Phys. Rev. Lett.}}\
  }%
  \textbf{\bibinfo {volume} {94}},\ \bibinfo {pages} {111801} (\bibinfo {year}
  {2005}),\ \Eprint{http://arxiv.org/abs/hep-ph/0405067}{arXiv:hep-ph/0405067
  [hep-ph]}%
  \bibAnnoteFile{NoStop}{Chacko:2004cz}%
\bibitem{Dasgupta:2013zpn}%
  \BibitemOpen
  \bibfield{author}{%
  \bibinfo {author} {\bibfnamefont{B.}~\bibnamefont{Dasgupta}}\ and\ \bibinfo
  {author} {\bibfnamefont{J.}~\bibnamefont{Kopp}},\ }%
  \bibfield{journal}{%
  \Doi{10.1103/PhysRevLett.112.031803}{\bibinfo {journal} {Phys. Rev. Lett.}}\
  }%
  \textbf{\bibinfo {volume} {112}},\ \bibinfo {pages} {031803} (\bibinfo {year}
  {2014}),\ \Eprint{http://arxiv.org/abs/1310.6337}{arXiv:1310.6337 [hep-ph]}%
  \bibAnnoteFile{NoStop}{Dasgupta:2013zpn}%
\bibitem{Mirizzi:2014ama}%
  \BibitemOpen
  \bibfield{author}{%
  \bibinfo {author} {\bibfnamefont{A.}~\bibnamefont{Mirizzi}}, \bibinfo
  {author} {\bibfnamefont{G.}~\bibnamefont{Mangano}}, \bibinfo {author}
  {\bibfnamefont{O.}~\bibnamefont{Pisanti}},\ and\ \bibinfo {author}
  {\bibfnamefont{N.}~\bibnamefont{Saviano}},\ }%
  \bibfield{journal}{%
  \Doi{10.1103/PhysRevD.91.025019}{\bibinfo {journal} {Phys. Rev.}}\ }%
  \textbf{\bibinfo {volume} {D91}},\ \bibinfo {pages} {025019} (\bibinfo {year}
  {2015}),\ \Eprint{http://arxiv.org/abs/1410.1385}{arXiv:1410.1385 [hep-ph]}%
  \bibAnnoteFile{NoStop}{Mirizzi:2014ama}%
\bibitem{Cherry:2014xra}%
  \BibitemOpen
  \bibfield{author}{%
  \bibinfo {author} {\bibfnamefont{J.~F.}\ \bibnamefont{Cherry}}, \bibinfo
  {author} {\bibfnamefont{A.}~\bibnamefont{Friedland}},\ and\ \bibinfo {author}
  {\bibfnamefont{I.~M.}\ \bibnamefont{Shoemaker}}}%
   (\bibinfo {year} {2014}),\
  \Eprint{http://arxiv.org/abs/1411.1071}{arXiv:1411.1071 [hep-ph]}%
  \bibAnnoteFile{NoStop}{Cherry:2014xra}%
\bibitem{Chu:2015ipa}%
  \BibitemOpen
  \bibfield{author}{%
  \bibinfo {author} {\bibfnamefont{X.}~\bibnamefont{Chu}}, \bibinfo {author}
  {\bibfnamefont{B.}~\bibnamefont{Dasgupta}},\ and\ \bibinfo {author}
  {\bibfnamefont{J.}~\bibnamefont{Kopp}},\ }%
  \bibfield{journal}{%
  \Doi{10.1088/1475-7516/2015/10/011}{\bibinfo {journal} {JCAP}}\ }%
  \textbf{\bibinfo {volume} {1510}},\ \bibinfo {pages} {011} (\bibinfo {year}
  {2015}),\ \Eprint{http://arxiv.org/abs/1505.02795}{arXiv:1505.02795
  [hep-ph]}%
  \bibAnnoteFile{NoStop}{Chu:2015ipa}%
\bibitem{Cherry:2016jol}%
  \BibitemOpen
  \bibfield{author}{%
  \bibinfo {author} {\bibfnamefont{J.~F.}\ \bibnamefont{Cherry}}, \bibinfo
  {author} {\bibfnamefont{A.}~\bibnamefont{Friedland}},\ and\ \bibinfo {author}
  {\bibfnamefont{I.~M.}\ \bibnamefont{Shoemaker}}}%
   (\bibinfo {year} {2016}),\
  \Eprint{http://arxiv.org/abs/1605.06506}{arXiv:1605.06506 [hep-ph]}%
  \bibAnnoteFile{NoStop}{Cherry:2016jol}%
\bibitem{Hochberg:2015pha}%
  \BibitemOpen
  \bibfield{author}{%
  \bibinfo {author} {\bibfnamefont{Y.}~\bibnamefont{Hochberg}}, \bibinfo
  {author} {\bibfnamefont{Y.}~\bibnamefont{Zhao}},\ and\ \bibinfo {author}
  {\bibfnamefont{K.~M.}\ \bibnamefont{Zurek}},\ }%
  \bibfield{journal}{%
  \Doi{10.1103/PhysRevLett.116.011301}{\bibinfo {journal} {Phys. Rev. Lett.}}\
  }%
  \textbf{\bibinfo {volume} {116}},\ \bibinfo {pages} {011301} (\bibinfo {year}
  {2016}),\ \Eprint{http://arxiv.org/abs/1504.07237}{arXiv:1504.07237
  [hep-ph]}%
  \bibAnnoteFile{NoStop}{Hochberg:2015pha}%
\bibitem{Hochberg:2015fth}%
  \BibitemOpen
  \bibfield{author}{%
  \bibinfo {author} {\bibfnamefont{Y.}~\bibnamefont{Hochberg}}, \bibinfo
  {author} {\bibfnamefont{M.}~\bibnamefont{Pyle}}, \bibinfo {author}
  {\bibfnamefont{Y.}~\bibnamefont{Zhao}},\ and\ \bibinfo {author}
  {\bibfnamefont{K.~M.}\ \bibnamefont{Zurek}},\ }%
  \bibfield{journal}{%
  \Doi{10.1007/JHEP08(2016)057}{\bibinfo {journal} {JHEP}}\ }%
  \textbf{\bibinfo {volume} {08}},\ \bibinfo {pages} {057} (\bibinfo {year}
  {2016}),\ \Eprint{http://arxiv.org/abs/1512.04533}{arXiv:1512.04533
  [hep-ph]}%
  \bibAnnoteFile{NoStop}{Hochberg:2015fth}%
\bibitem{Schutz:2016tid}%
  \BibitemOpen
  \bibfield{author}{%
  \bibinfo {author} {\bibfnamefont{K.}~\bibnamefont{Schutz}}\ and\ \bibinfo
  {author} {\bibfnamefont{K.~M.}\ \bibnamefont{Zurek}},\ }%
  \bibfield{journal}{%
  \Doi{10.1103/PhysRevLett.117.121302}{\bibinfo {journal} {Phys. Rev. Lett.}}\
  }%
  \textbf{\bibinfo {volume} {117}},\ \bibinfo {pages} {121302} (\bibinfo {year}
  {2016}),\ \Eprint{http://arxiv.org/abs/1604.08206}{arXiv:1604.08206
  [hep-ph]}%
  \bibAnnoteFile{NoStop}{Schutz:2016tid}%
\bibitem{Knapen:2016cue}%
  \BibitemOpen
  \bibfield{author}{%
  \bibinfo {author} {\bibfnamefont{S.}~\bibnamefont{Knapen}}, \bibinfo {author}
  {\bibfnamefont{T.}~\bibnamefont{Lin}},\ and\ \bibinfo {author}
  {\bibfnamefont{K.~M.}\ \bibnamefont{Zurek}},\ }%
  \bibfield{journal}{%
  \Doi{10.1103/PhysRevD.95.056019}{\bibinfo {journal} {Phys. Rev.}}\ }%
  \textbf{\bibinfo {volume} {D95}},\ \bibinfo {pages} {056019} (\bibinfo {year}
  {2017}),\ \Eprint{http://arxiv.org/abs/1611.06228}{arXiv:1611.06228
  [hep-ph]}%
  \bibAnnoteFile{NoStop}{Knapen:2016cue}%
\bibitem{Hut:1977zn}%
  \BibitemOpen
  \bibfield{author}{%
  \bibinfo {author} {\bibfnamefont{P.}~\bibnamefont{Hut}},\ }%
  \bibfield{journal}{%
  \Doi{10.1016/0370-2693(77)90139-3}{\bibinfo {journal} {Phys. Lett.}}\ }%
  \textbf{\bibinfo {volume} {B69}},\ \bibinfo {pages} {85} (\bibinfo {year}
  {1977})%
  \bibAnnoteFile{NoStop}{Hut:1977zn}%
\bibitem{Lee:1977ua}%
  \BibitemOpen
  \bibfield{author}{%
  \bibinfo {author} {\bibfnamefont{B.~W.}\ \bibnamefont{Lee}}\ and\ \bibinfo
  {author} {\bibfnamefont{S.}~\bibnamefont{Weinberg}},\ }%
  \bibfield{journal}{%
  \Doi{10.1103/PhysRevLett.39.165}{\bibinfo {journal} {Phys. Rev. Lett.}}\ }%
  \textbf{\bibinfo {volume} {39}},\ \bibinfo {pages} {165} (\bibinfo {year}
  {1977})%
  \bibAnnoteFile{NoStop}{Lee:1977ua}%
\bibitem{Enqvist:1991gx}%
  \BibitemOpen
  \bibfield{author}{%
  \bibinfo {author} {\bibfnamefont{K.}~\bibnamefont{Enqvist}}, \bibinfo
  {author} {\bibfnamefont{K.}~\bibnamefont{Kainulainen}},\ and\ \bibinfo
  {author} {\bibfnamefont{V.}~\bibnamefont{Semikoz}},\ }%
  \bibfield{journal}{%
  \Doi{10.1016/0550-3213(92)90359-J}{\bibinfo {journal} {Nucl. Phys.}}\ }%
  \textbf{\bibinfo {volume} {B374}},\ \bibinfo {pages} {392} (\bibinfo {year}
  {1992})%
  \bibAnnoteFile{NoStop}{Enqvist:1991gx}%
\bibitem{Pospelov:2007mp}%
  \BibitemOpen
  \bibfield{author}{%
  \bibinfo {author} {\bibfnamefont{M.}~\bibnamefont{Pospelov}}, \bibinfo
  {author} {\bibfnamefont{A.}~\bibnamefont{Ritz}},\ and\ \bibinfo {author}
  {\bibfnamefont{M.~B.}\ \bibnamefont{Voloshin}},\ }%
  \bibfield{journal}{%
  \Doi{10.1016/j.physletb.2008.02.052}{\bibinfo {journal} {Phys. Lett.}}\ }%
  \textbf{\bibinfo {volume} {B662}},\ \bibinfo {pages} {53} (\bibinfo {year}
  {2008}),\ \Eprint{http://arxiv.org/abs/0711.4866}{arXiv:0711.4866 [hep-ph]}%
  \bibAnnoteFile{NoStop}{Pospelov:2007mp}%
\bibitem{Raffelt:1996wa}%
  \BibitemOpen
  \bibfield{author}{%
  \bibinfo {author} {\bibfnamefont{G.~G.}\ \bibnamefont{Raffelt}},\ }%
  \emph{\bibinfo {title} {{Stars as laboratories for fundamental physics}}}\
  (\bibinfo {year} {1996})\ ISBN \bibinfo {isbn} {9780226702728},\
  \url{http://wwwth.mpp.mpg.de/members/raffelt/mypapers/199613.pdf}%
  \bibAnnoteFile{NoStop}{Raffelt:1996wa}%
\bibitem{Ade:2015xua}%
  \BibitemOpen
  \bibfield{author}{%
  \bibinfo {author} {\bibfnamefont{P.~A.~R.}\ \bibnamefont{Ade}} \emph{et~al.}
  (\bibinfo {collaboration} {Planck}),\ }%
  \bibfield{journal}{%
  \Doi{10.1051/0004-6361/201525830}{\bibinfo {journal} {Astron. Astrophys.}}\
  }%
  \textbf{\bibinfo {volume} {594}},\ \bibinfo {pages} {A13} (\bibinfo {year}
  {2016}),\ \Eprint{http://arxiv.org/abs/1502.01589}{arXiv:1502.01589
  [astro-ph.CO]}%
  \bibAnnoteFile{NoStop}{Ade:2015xua}%
\bibitem{Cyburt:2015mya}%
  \BibitemOpen
  \bibfield{author}{%
  \bibinfo {author} {\bibfnamefont{R.~H.}\ \bibnamefont{Cyburt}}, \bibinfo
  {author} {\bibfnamefont{B.~D.}\ \bibnamefont{Fields}}, \bibinfo {author}
  {\bibfnamefont{K.~A.}\ \bibnamefont{Olive}},\ and\ \bibinfo {author}
  {\bibfnamefont{T.-H.}\ \bibnamefont{Yeh}},\ }%
  \bibfield{journal}{%
  \Doi{10.1103/RevModPhys.88.015004}{\bibinfo {journal} {Rev. Mod. Phys.}}\ }%
  \textbf{\bibinfo {volume} {88}},\ \bibinfo {pages} {015004} (\bibinfo {year}
  {2016}),\ \Eprint{http://arxiv.org/abs/1505.01076}{arXiv:1505.01076
  [astro-ph.CO]}%
  \bibAnnoteFile{NoStop}{Cyburt:2015mya}%
\bibitem{Abazajian:2016yjj}%
  \BibitemOpen
  \bibfield{author}{%
  \bibinfo {author} {\bibfnamefont{K.~N.}\ \bibnamefont{Abazajian}}
  \emph{et~al.} (\bibinfo {collaboration} {CMB-S4})}%
   (\bibinfo {year} {2016}),\
  \Eprint{http://arxiv.org/abs/1610.02743}{arXiv:1610.02743 [astro-ph.CO]}%
  \bibAnnoteFile{NoStop}{Abazajian:2016yjj}%
\bibitem{Feng:2008mu}%
  \BibitemOpen
  \bibfield{author}{%
  \bibinfo {author} {\bibfnamefont{J.~L.}\ \bibnamefont{Feng}}, \bibinfo
  {author} {\bibfnamefont{H.}~\bibnamefont{Tu}},\ and\ \bibinfo {author}
  {\bibfnamefont{H.-B.}\ \bibnamefont{Yu}},\ }%
  \bibfield{journal}{%
  \Doi{10.1088/1475-7516/2008/10/043}{\bibinfo {journal} {JCAP}}\ }%
  \textbf{\bibinfo {volume} {0810}},\ \bibinfo {pages} {043} (\bibinfo {year}
  {2008}),\ \Eprint{http://arxiv.org/abs/0808.2318}{arXiv:0808.2318 [hep-ph]}%
  \bibAnnoteFile{NoStop}{Feng:2008mu}%
\bibitem{Berlin:2016gtr}%
  \BibitemOpen
  \bibfield{author}{%
  \bibinfo {author} {\bibfnamefont{A.}~\bibnamefont{Berlin}}, \bibinfo {author}
  {\bibfnamefont{D.}~\bibnamefont{Hooper}},\ and\ \bibinfo {author}
  {\bibfnamefont{G.}~\bibnamefont{Krnjaic}},\ }%
  \bibfield{journal}{%
  \Doi{10.1103/PhysRevD.94.095019}{\bibinfo {journal} {Phys. Rev.}}\ }%
  \textbf{\bibinfo {volume} {D94}},\ \bibinfo {pages} {095019} (\bibinfo {year}
  {2016}),\ \Eprint{http://arxiv.org/abs/1609.02555}{arXiv:1609.02555
  [hep-ph]}%
  \bibAnnoteFile{NoStop}{Berlin:2016gtr}%
\bibitem{Mangano:2001iu}%
  \BibitemOpen
  \bibfield{author}{%
  \bibinfo {author} {\bibfnamefont{G.}~\bibnamefont{Mangano}}, \bibinfo
  {author} {\bibfnamefont{G.}~\bibnamefont{Miele}}, \bibinfo {author}
  {\bibfnamefont{S.}~\bibnamefont{Pastor}},\ and\ \bibinfo {author}
  {\bibfnamefont{M.}~\bibnamefont{Peloso}},\ }%
  \bibfield{journal}{%
  \Doi{10.1016/S0370-2693(02)01622-2}{\bibinfo {journal} {Phys. Lett.}}\ }%
  \textbf{\bibinfo {volume} {B534}},\ \bibinfo {pages} {8} (\bibinfo {year}
  {2002}),\
  \Eprint{http://arxiv.org/abs/astro-ph/0111408}{arXiv:astro-ph/0111408
  [astro-ph]}%
  \bibAnnoteFile{NoStop}{Mangano:2001iu}%
\bibitem{Mangano:2005cc}%
  \BibitemOpen
  \bibfield{author}{%
  \bibinfo {author} {\bibfnamefont{G.}~\bibnamefont{Mangano}}, \bibinfo
  {author} {\bibfnamefont{G.}~\bibnamefont{Miele}}, \bibinfo {author}
  {\bibfnamefont{S.}~\bibnamefont{Pastor}}, \bibinfo {author}
  {\bibfnamefont{T.}~\bibnamefont{Pinto}}, \bibinfo {author}
  {\bibfnamefont{O.}~\bibnamefont{Pisanti}},\ and\ \bibinfo {author}
  {\bibfnamefont{P.~D.}\ \bibnamefont{Serpico}},\ }%
  \bibfield{journal}{%
  \Doi{10.1016/j.nuclphysb.2005.09.041}{\bibinfo {journal} {Nucl. Phys.}}\ }%
  \textbf{\bibinfo {volume} {B729}},\ \bibinfo {pages} {221} (\bibinfo {year}
  {2005}),\ \Eprint{http://arxiv.org/abs/hep-ph/0506164}{arXiv:hep-ph/0506164
  [hep-ph]}%
  \bibAnnoteFile{NoStop}{Mangano:2005cc}%
\bibitem{Riess:2016jrr}%
  \BibitemOpen
  \bibfield{author}{%
  \bibinfo {author} {\bibfnamefont{A.~G.}\ \bibnamefont{Riess}} \emph{et~al.},\
  }%
  \bibfield{journal}{%
  \Doi{10.3847/0004-637X/826/1/56}{\bibinfo {journal} {Astrophys. J.}}\ }%
  \textbf{\bibinfo {volume} {826}},\ \bibinfo {pages} {56} (\bibinfo {year}
  {2016}),\ \Eprint{http://arxiv.org/abs/1604.01424}{arXiv:1604.01424
  [astro-ph.CO]}%
  \bibAnnoteFile{NoStop}{Riess:2016jrr}%
\bibitem{Brust:2017nmv}%
  \BibitemOpen
  \bibfield{author}{%
  \bibinfo {author} {\bibfnamefont{C.}~\bibnamefont{Brust}}, \bibinfo {author}
  {\bibfnamefont{Y.}~\bibnamefont{Cui}},\ and\ \bibinfo {author}
  {\bibfnamefont{K.}~\bibnamefont{Sigurdson}}}%
   (\bibinfo {year} {2017}),\
  \Eprint{http://arxiv.org/abs/1703.10732}{arXiv:1703.10732 [astro-ph.CO]}%
  \bibAnnoteFile{NoStop}{Brust:2017nmv}%
\bibitem{Minkowski:1977sc}%
  \BibitemOpen
  \bibfield{author}{%
  \bibinfo {author} {\bibfnamefont{P.}~\bibnamefont{Minkowski}},\ }%
  \bibfield{journal}{%
  \Doi{10.1016/0370-2693(77)90435-X}{\bibinfo {journal} {Phys. Lett.}}\ }%
  \textbf{\bibinfo {volume} {B67}},\ \bibinfo {pages} {421} (\bibinfo {year}
  {1977})%
  \bibAnnoteFile{NoStop}{Minkowski:1977sc}%
\bibitem{Yanagida:1979as}%
  \BibitemOpen
  \bibfield{author}{%
  \bibinfo {author} {\bibfnamefont{T.}~\bibnamefont{Yanagida}},\ }%
  \bibfield{booktitle}{%
  \emph{\bibinfo {booktitle} {{Proceedings: Workshop on the Unified Theories
  and the Baryon Number in the Universe: Tsukuba, Japan, February 13-14,
  1979}}},\ }%
  \bibfield{journal}{%
  \bibinfo {journal} {Conf. Proc.}\ }%
  \textbf{\bibinfo {volume} {C7902131}},\ \bibinfo {pages} {95} (\bibinfo
  {year} {1979})%
  \bibAnnoteFile{NoStop}{Yanagida:1979as}%
\bibitem{Mohapatra:1979ia}%
  \BibitemOpen
  \bibfield{author}{%
  \bibinfo {author} {\bibfnamefont{R.~N.}\ \bibnamefont{Mohapatra}}\ and\
  \bibinfo {author} {\bibfnamefont{G.}~\bibnamefont{Senjanovic}},\ }%
  \bibfield{journal}{%
  \Doi{10.1103/PhysRevLett.44.912}{\bibinfo {journal} {Phys. Rev. Lett.}}\ }%
  \textbf{\bibinfo {volume} {44}},\ \bibinfo {pages} {912} (\bibinfo {year}
  {1980})%
  \bibAnnoteFile{NoStop}{Mohapatra:1979ia}%
\bibitem{GellMann:1980vs}%
  \BibitemOpen
  \bibfield{author}{%
  \bibinfo {author} {\bibfnamefont{M.}~\bibnamefont{Gell-Mann}}, \bibinfo
  {author} {\bibfnamefont{P.}~\bibnamefont{Ramond}},\ and\ \bibinfo {author}
  {\bibfnamefont{R.}~\bibnamefont{Slansky}},\ }%
  \bibfield{booktitle}{%
  \emph{\bibinfo {booktitle} {{Supergravity Workshop Stony Brook, New York,
  September 27-28, 1979}}},\ }%
  \bibfield{journal}{%
  \bibinfo {journal} {Conf. Proc.}\ }%
  \textbf{\bibinfo {volume} {C790927}},\ \bibinfo {pages} {315} (\bibinfo
  {year} {1979}),\ \Eprint{http://arxiv.org/abs/1306.4669}{arXiv:1306.4669
  [hep-th]}%
  \bibAnnoteFile{NoStop}{GellMann:1980vs}%
\bibitem{Schechter:1980gr}%
  \BibitemOpen
  \bibfield{author}{%
  \bibinfo {author} {\bibfnamefont{J.}~\bibnamefont{Schechter}}\ and\ \bibinfo
  {author} {\bibfnamefont{J.~W.~F.}\ \bibnamefont{Valle}},\ }%
  \bibfield{journal}{%
  \Doi{10.1103/PhysRevD.22.2227}{\bibinfo {journal} {Phys. Rev.}}\ }%
  \textbf{\bibinfo {volume} {D22}},\ \bibinfo {pages} {2227} (\bibinfo {year}
  {1980})%
  \bibAnnoteFile{NoStop}{Schechter:1980gr}%
\bibitem{Chikashige:1980ui}%
  \BibitemOpen
  \bibfield{author}{%
  \bibinfo {author} {\bibfnamefont{Y.}~\bibnamefont{Chikashige}}, \bibinfo
  {author} {\bibfnamefont{R.~N.}\ \bibnamefont{Mohapatra}},\ and\ \bibinfo
  {author} {\bibfnamefont{R.~D.}\ \bibnamefont{Peccei}},\ }%
  \bibfield{journal}{%
  \Doi{10.1016/0370-2693(81)90011-3}{\bibinfo {journal} {Phys. Lett.}}\ }%
  \textbf{\bibinfo {volume} {B98}},\ \bibinfo {pages} {265} (\bibinfo {year}
  {1981})%
  \bibAnnoteFile{NoStop}{Chikashige:1980ui}%
\bibitem{Gelmini:1980re}%
  \BibitemOpen
  \bibfield{author}{%
  \bibinfo {author} {\bibfnamefont{G.~B.}\ \bibnamefont{Gelmini}}\ and\
  \bibinfo {author} {\bibfnamefont{M.}~\bibnamefont{Roncadelli}},\ }%
  \bibfield{journal}{%
  \Doi{10.1016/0370-2693(81)90559-1}{\bibinfo {journal} {Phys. Lett.}}\ }%
  \textbf{\bibinfo {volume} {99B}},\ \bibinfo {pages} {411} (\bibinfo {year}
  {1981})%
  \bibAnnoteFile{NoStop}{Gelmini:1980re}%
\bibitem{Georgi:1981pg}%
  \BibitemOpen
  \bibfield{author}{%
  \bibinfo {author} {\bibfnamefont{H.~M.}\ \bibnamefont{Georgi}}, \bibinfo
  {author} {\bibfnamefont{S.~L.}\ \bibnamefont{Glashow}},\ and\ \bibinfo
  {author} {\bibfnamefont{S.}~\bibnamefont{Nussinov}},\ }%
  \bibfield{journal}{%
  \Doi{10.1016/0550-3213(81)90336-9}{\bibinfo {journal} {Nucl. Phys.}}\ }%
  \textbf{\bibinfo {volume} {B193}},\ \bibinfo {pages} {297} (\bibinfo {year}
  {1981})%
  \bibAnnoteFile{NoStop}{Georgi:1981pg}%
\bibitem{Aulakh:1982yn}%
  \BibitemOpen
  \bibfield{author}{%
  \bibinfo {author} {\bibfnamefont{C.~S.}\ \bibnamefont{Aulakh}}\ and\ \bibinfo
  {author} {\bibfnamefont{R.~N.}\ \bibnamefont{Mohapatra}},\ }%
  \bibfield{journal}{%
  \Doi{10.1016/0370-2693(82)90262-3}{\bibinfo {journal} {Phys. Lett.}}\ }%
  \textbf{\bibinfo {volume} {B119}},\ \bibinfo {pages} {136} (\bibinfo {year}
  {1982})%
  \bibAnnoteFile{NoStop}{Aulakh:1982yn}%
\bibitem{Bertolini:1987kz}%
  \BibitemOpen
  \bibfield{author}{%
  \bibinfo {author} {\bibfnamefont{S.}~\bibnamefont{Bertolini}}\ and\ \bibinfo
  {author} {\bibfnamefont{A.}~\bibnamefont{Santamaria}},\ }%
  \bibfield{journal}{%
  \Doi{10.1016/0550-3213(88)90100-9}{\bibinfo {journal} {Nucl. Phys.}}\ }%
  \textbf{\bibinfo {volume} {B310}},\ \bibinfo {pages} {714} (\bibinfo {year}
  {1988})%
  \bibAnnoteFile{NoStop}{Bertolini:1987kz}%
\bibitem{Babu:1991we}%
  \BibitemOpen
  \bibfield{author}{%
  \bibinfo {author} {\bibfnamefont{K.~S.}\ \bibnamefont{Babu}}\ and\ \bibinfo
  {author} {\bibfnamefont{R.~N.}\ \bibnamefont{Mohapatra}},\ }%
  \bibfield{journal}{%
  \Doi{10.1103/PhysRevLett.67.1498}{\bibinfo {journal} {Phys. Rev. Lett.}}\ }%
  \textbf{\bibinfo {volume} {67}},\ \bibinfo {pages} {1498} (\bibinfo {year}
  {1991})%
  \bibAnnoteFile{NoStop}{Babu:1991we}%
\bibitem{Feng:2010zp}%
  \BibitemOpen
  \bibfield{author}{%
  \bibinfo {author} {\bibfnamefont{J.~L.}\ \bibnamefont{Feng}}, \bibinfo
  {author} {\bibfnamefont{M.}~\bibnamefont{Kaplinghat}},\ and\ \bibinfo
  {author} {\bibfnamefont{H.-B.}\ \bibnamefont{Yu}},\ }%
  \bibfield{journal}{%
  \Doi{10.1103/PhysRevD.82.083525}{\bibinfo {journal} {Phys. Rev.}}\ }%
  \textbf{\bibinfo {volume} {D82}},\ \bibinfo {pages} {083525} (\bibinfo {year}
  {2010}),\ \Eprint{http://arxiv.org/abs/1005.4678}{arXiv:1005.4678 [hep-ph]}%
  \bibAnnoteFile{NoStop}{Feng:2010zp}%
\bibitem{Viel:2013apy}%
  \BibitemOpen
  \bibfield{author}{%
  \bibinfo {author} {\bibfnamefont{M.}~\bibnamefont{Viel}}, \bibinfo {author}
  {\bibfnamefont{G.~D.}\ \bibnamefont{Becker}}, \bibinfo {author}
  {\bibfnamefont{J.~S.}\ \bibnamefont{Bolton}},\ and\ \bibinfo {author}
  {\bibfnamefont{M.~G.}\ \bibnamefont{Haehnelt}},\ }%
  \bibfield{journal}{%
  \Doi{10.1103/PhysRevD.88.043502}{\bibinfo {journal} {Phys. Rev.}}\ }%
  \textbf{\bibinfo {volume} {D88}},\ \bibinfo {pages} {043502} (\bibinfo {year}
  {2013}),\ \Eprint{http://arxiv.org/abs/1306.2314}{arXiv:1306.2314
  [astro-ph.CO]}%
  \bibAnnoteFile{NoStop}{Viel:2013apy}%
\bibitem{Baur:2015jsy}%
  \BibitemOpen
  \bibfield{author}{%
  \bibinfo {author} {\bibfnamefont{J.}~\bibnamefont{Baur}}, \bibinfo {author}
  {\bibfnamefont{N.}~\bibnamefont{Palanque-Delabrouille}}, \bibinfo {author}
  {\bibfnamefont{C.}~\bibnamefont{Y�che}}, \bibinfo {author}
  {\bibfnamefont{C.}~\bibnamefont{Magneville}},\ and\ \bibinfo {author}
  {\bibfnamefont{M.}~\bibnamefont{Viel}},\ }%
  \bibfield{booktitle}{%
  \emph{\bibinfo {booktitle} {{SDSS-IV Collaboration Meeting, July 20-23,
  2015}}},\ }%
  \bibfield{journal}{%
  \Doi{10.1088/1475-7516/2016/08/012}{\bibinfo {journal} {JCAP}}\ }%
  \textbf{\bibinfo {volume} {1608}},\ \bibinfo {pages} {012} (\bibinfo {year}
  {2016}),\ \Eprint{http://arxiv.org/abs/1512.01981}{arXiv:1512.01981
  [astro-ph.CO]}%
  \bibAnnoteFile{NoStop}{Baur:2015jsy}%
\bibitem{Abazajian:2017tcc}%
  \BibitemOpen
  \bibfield{author}{%
  \bibinfo {author} {\bibfnamefont{K.~N.}\ \bibnamefont{Abazajian}}}%
   (\bibinfo {year} {2017}),\
  \Eprint{http://arxiv.org/abs/1705.01837}{arXiv:1705.01837 [hep-ph]}%
  \bibAnnoteFile{NoStop}{Abazajian:2017tcc}%
\bibitem{Sitwell:2013fpa}%
  \BibitemOpen
  \bibfield{author}{%
  \bibinfo {author} {\bibfnamefont{M.}~\bibnamefont{Sitwell}}, \bibinfo
  {author} {\bibfnamefont{A.}~\bibnamefont{Mesinger}}, \bibinfo {author}
  {\bibfnamefont{Y.-Z.}\ \bibnamefont{Ma}},\ and\ \bibinfo {author}
  {\bibfnamefont{K.}~\bibnamefont{Sigurdson}},\ }%
  \bibfield{journal}{%
  \Doi{10.1093/mnras/stt2392}{\bibinfo {journal} {Mon. Not. Roy. Astron.
  Soc.}}\ }%
  \textbf{\bibinfo {volume} {438}},\ \bibinfo {pages} {2664} (\bibinfo {year}
  {2014}),\ \Eprint{http://arxiv.org/abs/1310.0029}{arXiv:1310.0029
  [astro-ph.CO]}%
  \bibAnnoteFile{NoStop}{Sitwell:2013fpa}%
\bibitem{Sekiguchi:2014wfa}%
  \BibitemOpen
  \bibfield{author}{%
  \bibinfo {author} {\bibfnamefont{T.}~\bibnamefont{Sekiguchi}}\ and\ \bibinfo
  {author} {\bibfnamefont{H.}~\bibnamefont{Tashiro}},\ }%
  \bibfield{journal}{%
  \Doi{10.1088/1475-7516/2014/08/007}{\bibinfo {journal} {JCAP}}\ }%
  \textbf{\bibinfo {volume} {1408}},\ \bibinfo {pages} {007} (\bibinfo {year}
  {2014}),\ \Eprint{http://arxiv.org/abs/1401.5563}{arXiv:1401.5563
  [astro-ph.CO]}%
  \bibAnnoteFile{NoStop}{Sekiguchi:2014wfa}%
\bibitem{Shimabukuro:2014ava}%
  \BibitemOpen
  \bibfield{author}{%
  \bibinfo {author} {\bibfnamefont{H.}~\bibnamefont{Shimabukuro}}, \bibinfo
  {author} {\bibfnamefont{K.}~\bibnamefont{Ichiki}}, \bibinfo {author}
  {\bibfnamefont{S.}~\bibnamefont{Inoue}},\ and\ \bibinfo {author}
  {\bibfnamefont{S.}~\bibnamefont{Yokoyama}},\ }%
  \bibfield{journal}{%
  \Doi{10.1103/PhysRevD.90.083003}{\bibinfo {journal} {Phys. Rev.}}\ }%
  \textbf{\bibinfo {volume} {D90}},\ \bibinfo {pages} {083003} (\bibinfo {year}
  {2014}),\ \Eprint{http://arxiv.org/abs/1403.1605}{arXiv:1403.1605
  [astro-ph.CO]}%
  \bibAnnoteFile{NoStop}{Shimabukuro:2014ava}%
\bibitem{Tulin:2013teo}%
  \BibitemOpen
  \bibfield{author}{%
  \bibinfo {author} {\bibfnamefont{S.}~\bibnamefont{Tulin}}, \bibinfo {author}
  {\bibfnamefont{H.-B.}\ \bibnamefont{Yu}},\ and\ \bibinfo {author}
  {\bibfnamefont{K.~M.}\ \bibnamefont{Zurek}},\ }%
  \bibfield{journal}{%
  \Doi{10.1103/PhysRevD.87.115007}{\bibinfo {journal} {Phys. Rev.}}\ }%
  \textbf{\bibinfo {volume} {D87}},\ \bibinfo {pages} {115007} (\bibinfo {year}
  {2013}),\ \Eprint{http://arxiv.org/abs/1302.3898}{arXiv:1302.3898 [hep-ph]}%
  \bibAnnoteFile{NoStop}{Tulin:2013teo}%
\bibitem{Tulin:2012wi}%
  \BibitemOpen
  \bibfield{author}{%
  \bibinfo {author} {\bibfnamefont{S.}~\bibnamefont{Tulin}}, \bibinfo {author}
  {\bibfnamefont{H.-B.}\ \bibnamefont{Yu}},\ and\ \bibinfo {author}
  {\bibfnamefont{K.~M.}\ \bibnamefont{Zurek}},\ }%
  \bibfield{journal}{%
  \Doi{10.1103/PhysRevLett.110.111301}{\bibinfo {journal} {Phys. Rev. Lett.}}\
  }%
  \textbf{\bibinfo {volume} {110}},\ \bibinfo {pages} {111301} (\bibinfo {year}
  {2013}),\ \Eprint{http://arxiv.org/abs/1210.0900}{arXiv:1210.0900 [hep-ph]}%
  \bibAnnoteFile{NoStop}{Tulin:2012wi}%
\bibitem{Cyr-Racine:2013jua}%
  \BibitemOpen
  \bibfield{author}{%
  \bibinfo {author} {\bibfnamefont{F.-Y.}\ \bibnamefont{Cyr-Racine}}\ and\
  \bibinfo {author} {\bibfnamefont{K.}~\bibnamefont{Sigurdson}},\ }%
  \bibfield{journal}{%
  \Doi{10.1103/PhysRevD.90.123533}{\bibinfo {journal} {Phys. Rev.}}\ }%
  \textbf{\bibinfo {volume} {D90}},\ \bibinfo {pages} {123533} (\bibinfo {year}
  {2014}),\ \Eprint{http://arxiv.org/abs/1306.1536}{arXiv:1306.1536
  [astro-ph.CO]}%
  \bibAnnoteFile{NoStop}{Cyr-Racine:2013jua}%
\bibitem{Lancaster:2017ksf}%
  \BibitemOpen
  \bibfield{author}{%
  \bibinfo {author} {\bibfnamefont{L.}~\bibnamefont{Lancaster}}, \bibinfo
  {author} {\bibfnamefont{F.-Y.}\ \bibnamefont{Cyr-Racine}}, \bibinfo {author}
  {\bibfnamefont{L.}~\bibnamefont{Knox}},\ and\ \bibinfo {author}
  {\bibfnamefont{Z.}~\bibnamefont{Pan}}}%
   (\bibinfo {year} {2017}),\
  \Eprint{http://arxiv.org/abs/1704.06657}{arXiv:1704.06657 [astro-ph.CO]}%
  \bibAnnoteFile{NoStop}{Lancaster:2017ksf}%
\bibitem{Mangano:2006mp}%
  \BibitemOpen
  \bibfield{author}{%
  \bibinfo {author} {\bibfnamefont{G.}~\bibnamefont{Mangano}}, \bibinfo
  {author} {\bibfnamefont{A.}~\bibnamefont{Melchiorri}}, \bibinfo {author}
  {\bibfnamefont{P.}~\bibnamefont{Serra}}, \bibinfo {author}
  {\bibfnamefont{A.}~\bibnamefont{Cooray}},\ and\ \bibinfo {author}
  {\bibfnamefont{M.}~\bibnamefont{Kamionkowski}},\ }%
  \bibfield{journal}{%
  \Doi{10.1103/PhysRevD.74.043517}{\bibinfo {journal} {Phys. Rev.}}\ }%
  \textbf{\bibinfo {volume} {D74}},\ \bibinfo {pages} {043517} (\bibinfo {year}
  {2006}),\
  \Eprint{http://arxiv.org/abs/astro-ph/0606190}{arXiv:astro-ph/0606190
  [astro-ph]}%
  \bibAnnoteFile{NoStop}{Mangano:2006mp}%
\bibitem{Serra:2009uu}%
  \BibitemOpen
  \bibfield{author}{%
  \bibinfo {author} {\bibfnamefont{P.}~\bibnamefont{Serra}}, \bibinfo {author}
  {\bibfnamefont{F.}~\bibnamefont{Zalamea}}, \bibinfo {author}
  {\bibfnamefont{A.}~\bibnamefont{Cooray}}, \bibinfo {author}
  {\bibfnamefont{G.}~\bibnamefont{Mangano}},\ and\ \bibinfo {author}
  {\bibfnamefont{A.}~\bibnamefont{Melchiorri}},\ }%
  \bibfield{journal}{%
  \Doi{10.1103/PhysRevD.81.043507}{\bibinfo {journal} {Phys. Rev.}}\ }%
  \textbf{\bibinfo {volume} {D81}},\ \bibinfo {pages} {043507} (\bibinfo {year}
  {2010}),\ \Eprint{http://arxiv.org/abs/0911.4411}{arXiv:0911.4411
  [astro-ph.CO]}%
  \bibAnnoteFile{NoStop}{Serra:2009uu}%
\bibitem{Wilkinson:2014ksa}%
  \BibitemOpen
  \bibfield{author}{%
  \bibinfo {author} {\bibfnamefont{R.~J.}\ \bibnamefont{Wilkinson}}, \bibinfo
  {author} {\bibfnamefont{C.}~\bibnamefont{Boehm}},\ and\ \bibinfo {author}
  {\bibfnamefont{J.}~\bibnamefont{Lesgourgues}},\ }%
  \bibfield{journal}{%
  \Doi{10.1088/1475-7516/2014/05/011}{\bibinfo {journal} {JCAP}}\ }%
  \textbf{\bibinfo {volume} {1405}},\ \bibinfo {pages} {011} (\bibinfo {year}
  {2014}),\ \Eprint{http://arxiv.org/abs/1401.7597}{arXiv:1401.7597
  [astro-ph.CO]}%
  \bibAnnoteFile{NoStop}{Wilkinson:2014ksa}%
\bibitem{Boehm:2000gq}%
  \BibitemOpen
  \bibfield{author}{%
  \bibinfo {author} {\bibfnamefont{C.}~\bibnamefont{Boehm}}, \bibinfo {author}
  {\bibfnamefont{P.}~\bibnamefont{Fayet}},\ and\ \bibinfo {author}
  {\bibfnamefont{R.}~\bibnamefont{Schaeffer}},\ }%
  \bibfield{journal}{%
  \Doi{10.1016/S0370-2693(01)01060-7}{\bibinfo {journal} {Phys. Lett.}}\ }%
  \textbf{\bibinfo {volume} {B518}},\ \bibinfo {pages} {8} (\bibinfo {year}
  {2001}),\
  \Eprint{http://arxiv.org/abs/astro-ph/0012504}{arXiv:astro-ph/0012504
  [astro-ph]}%
  \bibAnnoteFile{NoStop}{Boehm:2000gq}%
\bibitem{Boehm:2004th}%
  \BibitemOpen
  \bibfield{author}{%
  \bibinfo {author} {\bibfnamefont{C.}~\bibnamefont{Boehm}}\ and\ \bibinfo
  {author} {\bibfnamefont{R.}~\bibnamefont{Schaeffer}},\ }%
  \bibfield{journal}{%
  \Doi{10.1051/0004-6361:20042238}{\bibinfo {journal} {Astron. Astrophys.}}\ }%
  \textbf{\bibinfo {volume} {438}},\ \bibinfo {pages} {419} (\bibinfo {year}
  {2005}),\
  \Eprint{http://arxiv.org/abs/astro-ph/0410591}{arXiv:astro-ph/0410591
  [astro-ph]}%
  \bibAnnoteFile{NoStop}{Boehm:2004th}%
\bibitem{Boehm:2014vja}%
  \BibitemOpen
  \bibfield{author}{%
  \bibinfo {author} {\bibfnamefont{C.}~\bibnamefont{Boehm}}, \bibinfo {author}
  {\bibfnamefont{J.~A.}\ \bibnamefont{Schewtschenko}}, \bibinfo {author}
  {\bibfnamefont{R.~J.}\ \bibnamefont{Wilkinson}}, \bibinfo {author}
  {\bibfnamefont{C.~M.}\ \bibnamefont{Baugh}},\ and\ \bibinfo {author}
  {\bibfnamefont{S.}~\bibnamefont{Pascoli}},\ }%
  \bibfield{journal}{%
  \Doi{10.1093/mnrasl/slu115}{\bibinfo {journal} {Mon. Not. Roy. Astron.
  Soc.}}\ }%
  \textbf{\bibinfo {volume} {445}},\ \bibinfo {pages} {L31} (\bibinfo {year}
  {2014}),\ \Eprint{http://arxiv.org/abs/1404.7012}{arXiv:1404.7012
  [astro-ph.CO]}%
  \bibAnnoteFile{NoStop}{Boehm:2014vja}%
\bibitem{Schewtschenko:2015rno}%
  \BibitemOpen
  \bibfield{author}{%
  \bibinfo {author} {\bibfnamefont{J.~A.}\ \bibnamefont{Schewtschenko}},
  \bibinfo {author} {\bibfnamefont{C.~M.}\ \bibnamefont{Baugh}}, \bibinfo
  {author} {\bibfnamefont{R.~J.}\ \bibnamefont{Wilkinson}}, \bibinfo {author}
  {\bibfnamefont{C.}~\bibnamefont{B�hm}}, \bibinfo {author}
  {\bibfnamefont{S.}~\bibnamefont{Pascoli}},\ and\ \bibinfo {author}
  {\bibfnamefont{T.}~\bibnamefont{Sawala}},\ }%
  \bibfield{journal}{%
  \Doi{10.1093/mnras/stw1078}{\bibinfo {journal} {Mon. Not. Roy. Astron.
  Soc.}}\ }%
  \textbf{\bibinfo {volume} {461}},\ \bibinfo {pages} {2282} (\bibinfo {year}
  {2016}),\ \Eprint{http://arxiv.org/abs/1512.06774}{arXiv:1512.06774
  [astro-ph.CO]}%
  \bibAnnoteFile{NoStop}{Schewtschenko:2015rno}%
\bibitem{Boehm:2003xr}%
  \BibitemOpen
  \bibfield{author}{%
  \bibinfo {author} {\bibfnamefont{C.}~\bibnamefont{Boehm}}, \bibinfo {author}
  {\bibfnamefont{H.}~\bibnamefont{Mathis}}, \bibinfo {author}
  {\bibfnamefont{J.}~\bibnamefont{Devriendt}},\ and\ \bibinfo {author}
  {\bibfnamefont{J.}~\bibnamefont{Silk}},\ }%
  \bibfield{journal}{%
  \Doi{10.1111/j.1365-2966.2005.09032.x}{\bibinfo {journal} {Mon. Not. Roy.
  Astron. Soc.}}\ }%
  \textbf{\bibinfo {volume} {360}},\ \bibinfo {pages} {282} (\bibinfo {year}
  {2005}),\
  \Eprint{http://arxiv.org/abs/astro-ph/0309652}{arXiv:astro-ph/0309652
  [astro-ph]}%
  \bibAnnoteFile{NoStop}{Boehm:2003xr}%
\bibitem{Boehm:2001hm}%
  \BibitemOpen
  \bibfield{author}{%
  \bibinfo {author} {\bibfnamefont{C.}~\bibnamefont{Boehm}}, \bibinfo {author}
  {\bibfnamefont{A.}~\bibnamefont{Riazuelo}}, \bibinfo {author}
  {\bibfnamefont{S.~H.}\ \bibnamefont{Hansen}},\ and\ \bibinfo {author}
  {\bibfnamefont{R.}~\bibnamefont{Schaeffer}},\ }%
  \bibfield{journal}{%
  \Doi{10.1103/PhysRevD.66.083505}{\bibinfo {journal} {Phys. Rev.}}\ }%
  \textbf{\bibinfo {volume} {D66}},\ \bibinfo {pages} {083505} (\bibinfo {year}
  {2002}),\
  \Eprint{http://arxiv.org/abs/astro-ph/0112522}{arXiv:astro-ph/0112522
  [astro-ph]}%
  \bibAnnoteFile{NoStop}{Boehm:2001hm}%
\bibitem{Choi:1989hi}%
  \BibitemOpen
  \bibfield{author}{%
  \bibinfo {author} {\bibfnamefont{K.}~\bibnamefont{Choi}}\ and\ \bibinfo
  {author} {\bibfnamefont{A.}~\bibnamefont{Santamaria}},\ }%
  \bibfield{journal}{%
  \Doi{10.1103/PhysRevD.42.293}{\bibinfo {journal} {Phys. Rev.}}\ }%
  \textbf{\bibinfo {volume} {D42}},\ \bibinfo {pages} {293} (\bibinfo {year}
  {1990})%
  \bibAnnoteFile{NoStop}{Choi:1989hi}%
\bibitem{Kachelriess:2000qc}%
  \BibitemOpen
  \bibfield{author}{%
  \bibinfo {author} {\bibfnamefont{M.}~\bibnamefont{Kachelriess}}, \bibinfo
  {author} {\bibfnamefont{R.}~\bibnamefont{Tomas}},\ and\ \bibinfo {author}
  {\bibfnamefont{J.~W.~F.}\ \bibnamefont{Valle}},\ }%
  \bibfield{journal}{%
  \Doi{10.1103/PhysRevD.62.023004}{\bibinfo {journal} {Phys. Rev.}}\ }%
  \textbf{\bibinfo {volume} {D62}},\ \bibinfo {pages} {023004} (\bibinfo {year}
  {2000}),\ \Eprint{http://arxiv.org/abs/hep-ph/0001039}{arXiv:hep-ph/0001039
  [hep-ph]}%
  \bibAnnoteFile{NoStop}{Kachelriess:2000qc}%
\bibitem{Farzan:2002wx}%
  \BibitemOpen
  \bibfield{author}{%
  \bibinfo {author} {\bibfnamefont{Y.}~\bibnamefont{Farzan}},\ }%
  \bibfield{journal}{%
  \Doi{10.1103/PhysRevD.67.073015}{\bibinfo {journal} {Phys. Rev.}}\ }%
  \textbf{\bibinfo {volume} {D67}},\ \bibinfo {pages} {073015} (\bibinfo {year}
  {2003}),\ \Eprint{http://arxiv.org/abs/hep-ph/0211375}{arXiv:hep-ph/0211375
  [hep-ph]}%
  \bibAnnoteFile{NoStop}{Farzan:2002wx}%
\bibitem{Heurtier:2016otg}%
  \BibitemOpen
  \bibfield{author}{%
  \bibinfo {author} {\bibfnamefont{L.}~\bibnamefont{Heurtier}}\ and\ \bibinfo
  {author} {\bibfnamefont{Y.}~\bibnamefont{Zhang}},\ }%
  \bibfield{journal}{%
  \Doi{10.1088/1475-7516/2017/02/042}{\bibinfo {journal} {JCAP}}\ }%
  \textbf{\bibinfo {volume} {1702}},\ \bibinfo {pages} {042} (\bibinfo {year}
  {2017}),\ \Eprint{http://arxiv.org/abs/1609.05882}{arXiv:1609.05882
  [hep-ph]}%
  \bibAnnoteFile{NoStop}{Heurtier:2016otg}%
\bibitem{Alekhin:2015byh}%
  \BibitemOpen
  \bibfield{author}{%
  \bibinfo {author} {\bibfnamefont{S.}~\bibnamefont{Alekhin}} \emph{et~al.},\
  }%
  \bibfield{journal}{%
  \Doi{10.1088/0034-4885/79/12/124201}{\bibinfo {journal} {Rept. Prog. Phys.}}\
  }%
  \textbf{\bibinfo {volume} {79}},\ \bibinfo {pages} {124201} (\bibinfo {year}
  {2016}),\ \Eprint{http://arxiv.org/abs/1504.04855}{arXiv:1504.04855
  [hep-ph]}%
  \bibAnnoteFile{NoStop}{Alekhin:2015byh}%
\bibitem{Knapen:2017xzo}%
  \BibitemOpen
  \bibfield{author}{%
  \bibinfo {author} {\bibfnamefont{S.}~\bibnamefont{Knapen}}, \bibinfo {author}
  {\bibfnamefont{T.}~\bibnamefont{Lin}},\ and\ \bibinfo {author}
  {\bibfnamefont{K.~M.}\ \bibnamefont{Zurek}},\ }%
  \bibfield{journal}{%
  \Doi{10.1103/PhysRevD.96.115021}{\bibinfo {journal} {Phys. Rev.}}\ }%
  \textbf{\bibinfo {volume} {D96}},\ \bibinfo {pages} {115021} (\bibinfo {year}
  {2017}),\ \Eprint{http://arxiv.org/abs/1709.07882}{arXiv:1709.07882
  [hep-ph]}%
  \bibAnnoteFile{NoStop}{Knapen:2017xzo}%
\bibitem{Battaglieri:2017aum}%
  \BibitemOpen
  \bibfield{author}{%
  \bibinfo {author} {\bibfnamefont{M.}~\bibnamefont{Battaglieri}}
  \emph{et~al.}}%
   (\bibinfo {year} {2017}),\
  \Eprint{http://arxiv.org/abs/1707.04591}{arXiv:1707.04591 [hep-ph]}%
  \bibAnnoteFile{NoStop}{Battaglieri:2017aum}%
\bibitem{Beacom:2004yd}%
  \BibitemOpen
  \bibfield{author}{%
  \bibinfo {author} {\bibfnamefont{J.~F.}\ \bibnamefont{Beacom}}, \bibinfo
  {author} {\bibfnamefont{N.~F.}\ \bibnamefont{Bell}},\ and\ \bibinfo {author}
  {\bibfnamefont{S.}~\bibnamefont{Dodelson}},\ }%
  \bibfield{journal}{%
  \Doi{10.1103/PhysRevLett.93.121302}{\bibinfo {journal} {Phys. Rev. Lett.}}\
  }%
  \textbf{\bibinfo {volume} {93}},\ \bibinfo {pages} {121302} (\bibinfo {year}
  {2004}),\
  \Eprint{http://arxiv.org/abs/astro-ph/0404585}{arXiv:astro-ph/0404585
  [astro-ph]}%
  \bibAnnoteFile{NoStop}{Beacom:2004yd}%
\end{thebibliography}%

\end{document}